\documentclass[prb,twocolumn,preprintnumbers,amsmath,amssymb,superscriptaddress]{revtex4}
\usepackage{epsf}
\usepackage{graphicx}
\usepackage{bm}
\usepackage{amsmath}
\usepackage{color}
\usepackage{notes2bib}
\usepackage{epstopdf}
\usepackage{textcomp}

\bibnotesetup{
note-name = , use-sort-key = false }
\begin{document}

\title{Chiral spin ordering of electron gas in solids with broken time reversal symmetry} 

\author{K.~S.~Denisov}
\email{denisokonstantin@gmail.com} \affiliation{Ioffe Institute, 194021 St.Petersburg, Russia}
\affiliation{Lappeenranta-Lahti University of Technology, FI-53851 Lappeenranta, Finland}
\author{I.~V.~Rozhansky}
\affiliation{Ioffe Institute, 194021 St.Petersburg, Russia}
\affiliation{Lappeenranta-Lahti University of Technology, FI-53851 Lappeenranta, Finland}
\author{N.~S.~Averkiev}
\affiliation{Ioffe Institute, 194021 St.Petersburg, Russia}
\author{E.~L\"ahderanta}
\affiliation{Lappeenranta-Lahti University of Technology, FI-53851 Lappeenranta, Finland}

\begin{abstract}
In this work we manifest that an electrostatic disorder 
in conducting systems with broken time reversal symmetry universally leads to a chiral ordering of the electron gas
giving rise to skyrmion-like textures in spatial distribution of the electron spin density. 
We describe a microscopic mechanism underlying the formation of the equilibrium chiral spin textures 
in two-dimensional systems with 
spin-orbit interaction and exchange spin splitting. 
We have obtained analytical expressions for spin-density response functions and have analyzed both  local and non-local spin response to electrostatic perturbations for systems 
with parabolic-like and Dirac electron spectra.
With the proposed theory we come up with a concept of 
controlling spin chirality by electrical means. 
\end{abstract}

\maketitle
The concept of spin chirality constitutes a substantial part of modern condensed matter physics. 
It is widely applied for strongly correlated electron systems~\cite{Chir_Wilczek,chen2010local,batista2016frustration,kallin2016chiral} when interpreting 
the fractional statistics~\cite{Kalmeyer1987,yang1993possible,kitaev2006anyons}, or chiral spin liquids~\cite{Volovik2016,lee1992gauge,bauer2014chiral,MessioPRL2017} in terms of an effective gauge field. 
Remarkably, a finite spin chirality 
induces a gauge invariant magnetic flux, which is an experimentally observable quantity~\cite{Chir_Wilczek}.
It was shown that the chirality driven magnetic field 
affects electron transport in the very same way as the ordinary magnetic field does~\cite{Ye1999,chun2000magnetotransport} leading to the Hall response, 
the phenomenon currently referred as the topological Hall effect~\cite{BrunoDugaev,denisov2018general,QTHE}. 
Naturally, to get an experimental access to the variety of spin chirality driven phenomena an efficient tool for creating chiral spin ordrer is needed. 
One way towards this goal is to focus on materials 
possessing exotic spin textures, such as magnetic skyrmions~\cite{fert2017magnetic,wiesendanger2016nanoscale,soumyanarayanan2017tunable,nakajima2017skyrmion}, or merons~\cite{yu2018transformation}. 
Still, exploring the physical mechanisms behind the emergence of spin chirality in solids remains challenging and is of high fundamental interest.

In this Letter we show that in systems with broken time reversal symmetry ($\mathcal{T}$-symmetry)
a chiral spin order of electron gas is universally induced by an electrostatic disorder, 
which is an inherent property of any real solid.
We argue that 
numerous crystal imperfections, such as 
residue impurities or surface defects 
appear to be a source of local chiral spin ordering in the electron gas. 
This effect is more pronounced for an electron gas with stronger  spin-orbit interaction (SOI).
Naturally, various magnetic 
systems 
such as magnetic topological insulators (TI)~\cite{tokura2019magnetic,OkadaPRL2011,WeiPRL2013,chen2010massive}, 
Rashba magnetic layers~\cite{miron2010current,miron2011perpendicular,manchon2015new,zhou2018observation} or dilute magnetic semiconductors (DMS)~\cite{liu2008quantum,yu2010quantized,chernyshov2009evidence,novik2005band,jungwirth2014spin,li2013tailoring}
are in fact flooded by chiral spin textures pinned to structural defects. 
This effect opens up 
a novel concept of an experimental research of spin chirality 
driven phenomena. 
In our work we focus on 
two-dimensional degenerate electron gas (2DEG) with 
a spin-orbit interaction and an exchange spin splitting. 
We introduce an effective 'magnetic field' acting on an electron spin:
\begin{equation}
\label{eq_Bk}
    \boldsymbol{B}_k =\left(
    \lambda k \cos{\left(\chi \varphi_k+\gamma \right)},
    \lambda k \sin{\left(\chi \varphi_k + \gamma \right)},
    h
    \right),
\end{equation}
where
$\boldsymbol{k}=(k,\varphi_k)$ is a 2D momentum with magnitude $k$ and polar angle $\varphi_k$.  
The parameter $h>0$ describes the out-of-plane component leading to the carrier spin splitting at $k=0$, 
it is thus responsible for the violation of $\mathcal{T}$-symmetry. 
The in-plane components of $\boldsymbol{B}_k$ represent linear in $k$ terms due to SOI, $\lambda$ is the SOI coupling constant. 
The SOI parameters $\chi =\pm 1 $ (helicity) and $\gamma$ (vorticity) 
cover different types of the SOI interaction.

\begin{figure}
	\centering	
	\includegraphics[width=0.5\textwidth]{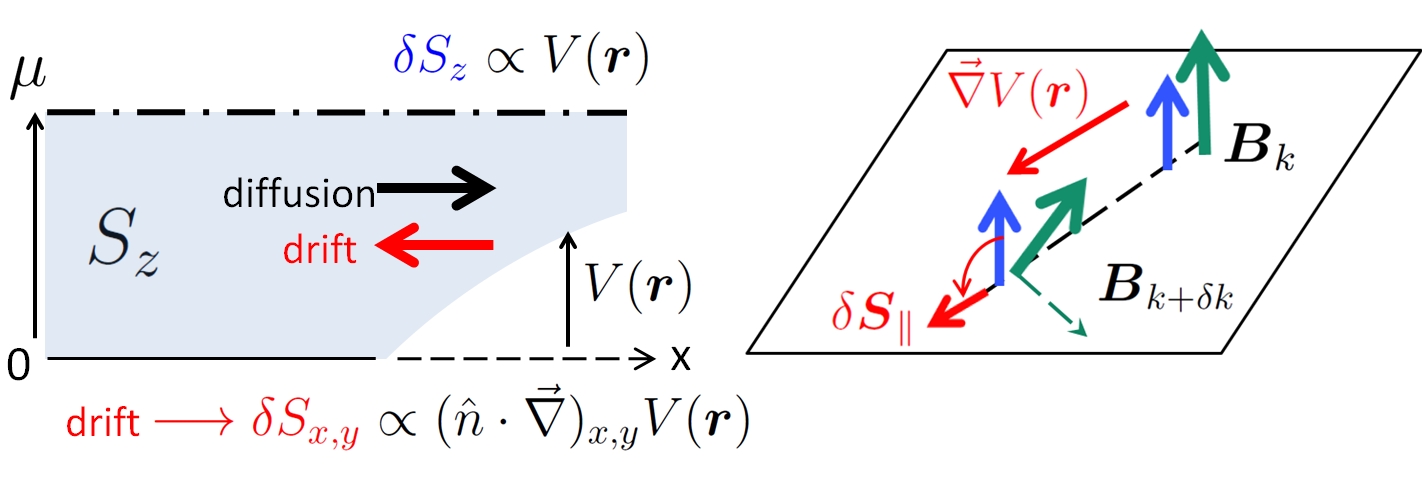}
	\caption{The physical picture behind the emergence of an equilibrium chiral spin pattern of the electron gas. The in-plane spin arises from the precession due to drift electron flow.}
	\label{f1}
\end{figure}

Let us further assume an electrostatic disorder due to various defects 
present in the system. 
When $\mathcal{T}$-symmetry is broken the 
spatial distribution of the equilibrium electron spin density 
follows the inhomogeneity of the electrostatic potential $V(\bm r)$. 
Indeed, at $h\neq 0$ there is a nonzero electron spin polarization directed perpendicular to the motion plane ($z$-axis). 
A local shift of $V(\bm r)$ 
leads to a spatial redistribution of electrons and, hence, to the change of $\delta S_z$. 
When SOI is present  ($\lambda \neq 0$) 
the in-plane components of the spin density $\delta S_{x,y}$ appear as well, so 
the induced spin response $\delta \bm S$ acquires a chiral spatial pattern forming skyrmion-like spin textures. 

Let us notice that the 
mechanics behind the appearance of $\delta S_{x,y}$ 
in response to an electrostatic potential
has a peculiar character, 
and it differs 
from that for $\delta S_z$. 
Since there is no net in-plane spin polarization at a spatially uniform electrostatic potential, 
 $\delta S_{x,y}$ appears only due to its gradient. 
One can consider the following quaisclassical picture, see Fig.\ref{f1}.
An electron with initial momentum $\bm k$ and spin $\bm S_{k}$ co-aligned 
with the direction of $\bm B_{k}$
moves along a certain trajectory.
Due to the electrostatic potential 
gradient $ \Vec{\nabla} V(\bm{r})$  
the carrier  momentum is changed $\bm k + \delta \bm k$,
thus changing the tilt of the magnetic field $\bm B_{k+\delta k}$,
which in-plane components are coupled with momentum. 
This process triggers the precession of electron spin 
around the new direction of the magnetic field
creating an excessive in-plane spin density.
In the thermodynamic equilibrium there is no net current
as the drift and diffusion electron flows are compensated everywhere.
However, the in-plane components of the spin density still appear 
because the drift flow is associated with a change of the electron momentum. 

The emergence of the spin textures due to spatial variation of the electrostatic 
potential is described by static spin-density response functions: 
\begin{align}
\label{eq_First}
    & \mathcal{F}_{\alpha}(\boldsymbol{q}) = \sum_{k,s,s'} \langle u_k^s| \hat{S}_{\alpha} | u_{k+q}^{s'} \rangle \langle u_{k+q}^{s'}| u_{k}^{s} \rangle  \frac{f_k^s - f_{k+q}^{s'}}{\varepsilon_k^s - \varepsilon_{k+q}^{s'}+i0}, 
\end{align}
where 
$\hat{S}_{\alpha}$ is a spin operator for $\alpha  = (x,y,z)$ axis, 
index $s=\pm$ denotes
two electron subbands, 
$\varepsilon_k^s$ and $| u_{k}^{s} \rangle $ are the energy and the Bloch amplitude of an electron in state $(\boldsymbol{k},s)$, 
 $f_k^s$ is  
the equilibrium distribution function.
Using the functions $\mathcal{F}_{\alpha}(\bm{q})$ one can analyze 
the  spin density $\delta \bm{S}(\bm{r})$ emerging in 2DEG in the 
vicinity of a doping center or a defect characterized by  a 
potential ${V}(\boldsymbol{r})$: 
\begin{align}
\label{eq_SparSz}
    & \delta \bm{S}_{\alpha}(\boldsymbol{r}) = \int \frac{d\boldsymbol{q}}{(2\pi)^2} e^{i \boldsymbol{qr}} \mathcal{F}_{\alpha}(\bm{q}) V(\boldsymbol{q}),
\end{align}
where $V(\boldsymbol{q})$ is the Fourier transform of $V(\boldsymbol{r})$, the electron-electron interaction is neglected. 
In particular, the functions $\mathcal{F}_{\alpha}(\bm q)$  
allow us to identify whether the spin response is  local  or 
 extends beyond the localization radius of the potential  
due to the wave properties of 2DEG.

Let us point out a few general features 
of the spin response in the considered model.
As has been mentioned  above, 
$z$-component of spin is analogous to the electron density, so 
if there is no spatial dispersion 
$\delta S_z$ locally couples with the potential 
$\delta S_z(\bm r) = \varkappa_z V(\bm r)$. 
The coefficient $\varkappa_z$ is given by 
a product of the electron density of states and $z$-projection of spin taken at the Fermi energy. 
On the contrary,   
the in-plane spin components driven by the precession mechanism illustrated by Fig.~\ref{f1} 
are induced by the gradient of $V(\bm r)$. 
In the case of a local response this coupling takes the form
$\delta S_{x,y} = \varkappa_\parallel ( \hat{n} \cdot \Vec{\nabla}  )_{x,y} V(\bm r)$, 
where $\hat{n}$ is a unitary matrix determined by SOI type,
and the coefficient $\varkappa_\parallel$ is determined by a carrier spectrum. 
Since the Fourier transform of $\Vec{\nabla} V(\bm r)$ is $
i \bm q V(\bm q)$, we conclude 
that $\mathcal{F}_{x,y}(\bm q)$ are purely imaginary, 
we present them as
\begin{equation}
    \mathcal{F}_{x,y}(\boldsymbol{q}) = i 
    \left( \hat{n} \bm e_q \right)_{x,y}
    \mathcal{F}_{\parallel}(q),
\end{equation}
where the real function $\mathcal{F}_\parallel(q)$ depends on the absolute value of $q$, 
$\bm e_q = \bm q /q$. 
Naturally, 
$\mathcal{F}_\parallel(q) \propto q$ at $q\to 0$. 

As soon as there is some spatial inhomogeneity of crystalline structure the electron gas acquires a local spin chirality. 
For an axially symmetric potential $V(r)$ 
the excessive 
spin density $\delta \boldsymbol{S}(\boldsymbol{r})$ 
profile has a shape:
\begin{align}
\label{eq_deltaSr}
&    \delta \boldsymbol{S}(\boldsymbol{r}) = \begin{pmatrix} 
    \delta S_{\parallel}(r) \cos{\left(\chi \varphi_r+\gamma' \right)} \\
    \delta S_{\parallel}(r) \sin{\left(\chi \varphi_r + \gamma' \right)}\\
    \delta S_{z}(r)
    \end{pmatrix},
    \\
& \delta S_{z,\parallel}(r) = \int \frac{qdq}{2\pi} J_{0,1}(qr) \mathcal{F}_{z,\parallel}(q) V(q), 
\notag
\end{align}
where $r,\varphi_r$ are the polar coordinates of a radius vector,
$\delta S_{\parallel}, \delta S_{z}$ depend on $r$,  $J_{0,1}$ are Bessel's functions of the zeroth and first order, respectively, $\gamma' = \gamma + \pi/2$. 
The emerging chiral spin cloud is similar to a skyrmion for $\chi=1$, or to an antiskyrmion for $\chi=-1$. 
The fact that the helicity $\gamma'$ of the real space spin rotation is shifted by $\pi/2$ with respect to $\gamma$ in $k$-space 
reflects the 
spin precession mechanism. 
The details of chiral spin response 
naturally depend on a carrier band structure $\varepsilon_k^\pm$. 
Below we calculate $\mathcal{F}_{z,\parallel}(q) = \mathcal{F}_{z,\parallel}^+ + \mathcal{F}_{z,\parallel}^-$, which is a sum of the responses of two subbands (see Supplementary materials), and 
analyze the spin response for parabolic-like and Dirac electron spectra.

\textit{Parabolic-like spectrum.}
Let us assume the following Hamiltonian ${H}_k$ and the energy spectrum $\varepsilon_k^s$:
\begin{align}
\label{eq_QW_spectr}
    & {H}_k = \frac{k^2}{2m} -\boldsymbol{B}_k \cdot \hat{\boldsymbol{\sigma}}, \\
    & \varepsilon_k^{\pm} = \frac{k^2}{2m} \mp  B_k, 
    \hspace{0.5cm}
    B_k = \sqrt{h^2 + (\lambda k)^2},
    \notag
\end{align}
where $m$ is the effective mass in the absence of the field $B_k$ (we assume $\hbar$=1). 
In this paper we take the parameter $\xi = m \lambda^2/h <1$. 
The spectrum of the system is shown in Fig.~\ref{f2}a, the color within each subband indicates 
the magnitude of $\zeta_{s} = \lambda k_{s}/h$, which has a meaning of 
spin inclination into the plane of the carrier motion
(blue color corresponds to $\zeta_{s} \ll 1$, red color 
indicates $\zeta_{s} \gg 1$).

We have obtained  analytic expressions for the spin-density response functions $\mathcal{F}_{z,\parallel}$ with the spectrum given by Eq.~\ref{eq_QW_spectr}. 
As the formulas are rather cumbersome, we provide them in the Supplementary Materials. 
Importantly, $\mathcal{F}_{z,\parallel}^{\pm}$ within each subband are decomposed onto a sum of intra- and interband contributions $\mathcal{F}_{z,\parallel}^{\pm} = \mathcal{F}_{z,\parallel}^{\pm \pm} +  \mathcal{F}_{z,\parallel}^{\pm \mp}$  
with the interband terms exhibiting an additional coupling $\mathcal{F}_z^{\pm \mp}(q) =  (2 m \lambda/q) \mathcal{F}_{\parallel}^{\pm \mp}(q)$.


Let us firstly consider the local coupling regime, when 
$\delta S_{x,y}(\bm r) = \varkappa_\parallel (\hat{n} \cdot \Vec{\nabla})_{x,y}  V(\bm r)$ and 
$\delta S_z(\bm r) = \varkappa_z V(\bm r)$. 
The coefficients $\varkappa_{z,\parallel} = \varkappa_{z,\parallel}^+ + \varkappa_{z,\parallel}^-$ 
 found from the limiting behavior of $\mathcal{F}_{z,\parallel}^\pm$ at 
 $q\to 0$ are given in Supplementary Materials. 
As already mentioned, $\varkappa_z^\pm$ is 
the product of the density of states 
and the spin $z$-projection at the Fermi energy $\mu$. 
The dependence of $\varkappa_{z,\parallel}$ on $\mu$ is shown in Fig~\ref{f2}b,c.
We note that nonzero spin response is observed only when 
the upper subband is free of electrons ($\mu<h$). 
This result is an inherent property of the considered model; the 
background spin density $S_z^{0} = (mh/4 \pi)$ remains constant at $\mu > h$. 

As follows from the 
explicit expressions for $\mathcal{F}_{z,\parallel}^\pm$,  
the local coupling regime occurs 
when the Fourier components of $V(\bm q)$ are localized within 
$q \ll {\rm min}[k_\pm, a_0^{-1}]$, where $a_0 = \lambda/2h$. 
For these values the response functions $\mathcal{F}_z$,
$\mathcal{F}_\parallel/q$
have a weak dependence on $q$, which 
means no spatial dispersion and, thus, the absence of non-locality in the response. 
Note, that, apart from the Fermi wavevector $k_\pm$, 
there is a second spatial scale $a_0 = \lambda/2h$, which 
controls the spatial dispersion of the spin response. 
This scale is associated with the precession 
mechanism for the in-plane spin generation. 

Naturally, a more interesting spin physics takes place when the effects of spatial dispersion come to the fore. 
We first consider the case when only the $(+)$ subband is populated ($ -h<\mu < h$).
The dependence of $\mathcal{F}_{z,\parallel}^{+}$, and its partial contributions $\mathcal{F}_{z,\parallel}^{+ \pm}$ on $q/2k_+$ 
are shown in Fig.~\ref{f3}(a,b). 
As we have discussed above, the response functions at $q \to 0$ behave as 
$\mathcal{F}_z^+\sim q^0$,
$\mathcal{F}_\parallel^+\sim q^1$. 
Another general trend is that the intra- $\mathcal{F}_{z,\parallel}^{++}$ and interband $\mathcal{F}_{z,\parallel}^{+-}$ terms
have an opposite sign and, thus, tend to cancel each other. 
The spatial profile $\delta S_{z,\parallel}(r)$ 
induced around a repulsive short range potential $V(r) = \alpha_0 \delta(\boldsymbol{r})$ is shown in Fig.~\ref{f3}c. 
The largest spin response appears 
within the Fermi wavelength ($2k_+ r \lesssim 2$). 
Going away from the center 
$S_{z,\parallel}(r)$ decrease exhibiting the Friedel oscillations with the period ${2k_+}$ (see  inset in Fig.\ref{f3}). 
Let us now  consider the case 
$\mu>h$ when both spin subbands are populated.  
Although the local spin response is absent in this case ($\varkappa_{z,\parallel}=0)$, 
the effect of spatial dispersion restores a chiral spin pattern. 
Shown in Fig.~\ref{f3}(d-g) are the calculated spin response functions $\mathcal{F}_{z,\parallel}^{\pm}(q)$.
We note that the intraband terms $\mathcal{F}_{z,\parallel}^{\pm \pm}$ 
for both subbands exhibit a single spike at $q=2k_\pm$. 
What is more interesting is the double-spike structure of the interband terms 
 $\mathcal{F}_{z,\parallel}^{\pm \mp}$ driven by 
 two nesting vectors connecting two distinct subbands of the Fermi surface. 
The presence of the two different spatial scales $k_\pm$ along with a complex structure of interband transitions 
lead to quite a peculiar spin response in the real space. 
In Fig.~\ref{f3}h,i we demonstrate $\delta S_{z,\parallel}(r)$ 
for the short range potential $V(q) = \alpha_0$. 
The spatial scale is given in units of $x= k_F r$, where $k_F$ is the averaged Fermi wavevector. 
The Friedel oscillations clearly visible at large distances from the centre $x\gg 1$ 
are now formed by the superposition of oscillations with different spatial periods.
\textit{Dirac spectrum.} 
Let us further consider the case of Hamiltonian with Dirac spectrum:
\begin{equation}
\label{eq_TI_spectr}
    {H}_k = -\boldsymbol{B}_k \cdot \hat{\boldsymbol{\sigma}}, 
    \hspace{0.3cm}
    \varepsilon_k^{\pm} = \mp \sqrt{h^2 + (\lambda k)^2}. 
\end{equation}
This model describes, for example, 
chiral surface states of a 3D TI (the SOI parameters are $\chi=1$, $\gamma = \pi$).
Shown in Fig.~\ref{f4}a is the spectrum (\ref{eq_TI_spectr}), which consists of two 
nearly linear bands separated by the gap $2h$. 
The fundamental difference 
from the previously considered parabolic-like spectrum 
is the additional electron-hole symmetry ($\mathcal{C}$-symmetry), 
which modifies the electron gas response to external perturbations
~\cite{GrapheneHwang,wunsch2006dynamical,ZhuPRL2011,checkelsky2012dirac}. 

Let us put the Fermi energy $\mu > 0$ above the charge neutrality point, 
so the lower $(+)$ band is completely filled while the upper $(-)$ band is filled partially. 
Our calculations show that 
the fillled lower subband does not contribute to the response of $z$ spin component ($\mathcal{F}_z^+ =0$), 
while 
for the upper $(-)$ subband 
the spin response function 
is the conventional 2D Lindhard function: 
\begin{align}
\label{eq_FTI}
& \mathcal{F}_{z}^-(q) = \frac{m_{g}}{4\pi} \left( 1 - \Theta[q-2k_-] \sqrt{1 - 4 k_-^2/q^2} \right), 
\end{align}
where $m_{g} = h/\lambda^2$ is an effective mass due to the spectrum gap, 
$k_- = \sqrt{\mu^2 - h^2}/\lambda$ is the Fermi wavevector. 
This result is rather interesting as 
the Lindhard function usually describes the susceptibility of a system with simple parabolic spectrum.

Another important feature of $\mathcal{F}_z^-$ given by (\ref{eq_FTI}) is that 
its magnitude does not depend on the Fermi energy $\mu$. 
This is in contrast with 
the parabolic-like case, where the increase of $\mu$ leads to the suppression of spin response according to $\mathcal{F}_z^\pm \propto 1/\zeta_\pm$ at $\zeta_\pm \gg 1$. 
This effect is due to 
the density of states, which for the Dirac spectrum takes the form $\nu_- = B_k/2\pi \lambda^2$. 
Upon the increase of the Fermi energy the suppression of spin $z$-projection (which is $h/2B_k \propto {1}/{\zeta_-}$ at $\mu \gg h$) 
is exacltly compensated by the increase of $\nu_-(\mu)$.  
For instance, considering the local coupling regime $\delta S_z = \varkappa_z V(\bm r)$ 
the spin response is explicitly determined by a product $\varkappa_z^- = \nu_-(\mu) \cdot n_k/2 = (m_g/4\pi)$ independent of $\mu$. 

The in-plane spin response also exhibits a number of peculiar features. 
For the  functions $\mathcal{F}_{\parallel}^{\pm}(q)$ we obtained:
\begin{align}
\label{eq_FTI-2}
& \mathcal{F}_{\parallel}^+(q) = \frac{m_{g}}{4\pi} \tan^{-1}\left(qa_0 \right) ,
\\
&\mathcal{F}_{\parallel}^-= - \mathcal{F}_{\parallel}^+
+ 
\frac{m_{g}}{4\pi} \Theta[q-2k_-] 
\tan^{-1}\left( a_0 \sqrt{ \frac{q^2 - 4k_-^2}{ 1 + \zeta_-^2} } \right).
\notag
\end{align}
We note that there is a non-zero spin response from the completely filled $(+)$ subband, 
and that $\mathcal{F}_\parallel^+/q$ remains finite 
even at $qa_0 \ll 1$. 
This is an unusual behavior, since no density response of $(+)$ subband can be induced in this case. 
Indeed, the interband transitions underlying the change of electron density are suppressed for a smooth potential 
$q a_0 \ll 1$ due to the finite band gap $2h$. 
On the contrary, the in-plane spin response originates from the spin precession driven by a drift electron flow, which 
remains finite in $\mathcal{C}$-symmetry systems even with gaped spectrum due to the Klein tunneling.

Considering the in-plane spin response from the upper $(-)$ subband we note that 
the function $\mathcal{F}_\parallel^-(q)$ given by Eq.~\ref{eq_FTI-2} 
contains both the $\mu$-independent term opposite to that of $(+)$ subband, 
and a $\mu$-aware contribution responsible for the Friedel's oscillations with the spatial period $2 k_-$. 
The in-plane spin response function $\mathcal{F}_\parallel = \mathcal{F}_\parallel^+ +  \mathcal{F}_\parallel^-$ and its partial components  
$\mathcal{F}_{\parallel}^\pm$ are shown in Fig.~\ref{f4}b.
The contributions of $(\pm)$ subbands cancel each other at $q<2k_-$ and  $\mathcal{F}_\parallel$ turns to zero. 
Therefore, no in-plane spin response is induced by a long-range electrostatic perturbation 
when the Fermi level is in the upper subband.

The non-local spin response is also modified due to $\mathcal{C}$-symmetry.
As can be seen in Fig.~\ref{f4}b the function $\mathcal{F}_\parallel(q) = \mathcal{F}_\parallel^++\mathcal{F}_\parallel^-$ 
saturates at $q \gg k_-$ instead of going to zero.
However, as discussed above, the in-plane spin density responds 
to the potential gradient, so it is $\mathcal{F}_\parallel(q)/q$ 
which has the physical meaning and it indeed decays as $1/q$ when 
$\mathcal{F}_\parallel(q)$ saturates.
In Fig.~\ref{f4}c we show 
the spatial spin pattern $\delta S_{z,\parallel}$ induced by a short-range potential $V(q) = \alpha_0 {\rm exp}[{-(q a/2)^2}]$, $a$ is a potential radius.
It is worth mentioning that the magnitude of the in-plane spin response $\delta S_\parallel$ in the vicinity of a defect is far larger than in the 
parabolic spectra case due to the saturation of $\mathcal{F}_\parallel(q)$.
This finding emphasizes a particularly high susceptibility of chiral spin pattern in response to an electrostatic disorder in systems with Dirac spectrum. 

\textit{Discussion.}
Our study suggests that the emergence of chiral spin textures 
driven by an electrostatic disorder is a universal phenomena. 
The obtained results are applicable
to a variety of 
experimentally studied systems, such as DMS~\cite{GajKos,jungwirth2014spin}, thin films of 
ferromagnets~\cite{miron2010current,miron2011perpendicular},
Bi$_2$Se$_3$ doped by 
magnetic impurities~\cite{WeiPRL2013,chen2010massive,kou2012magnetically,watson2013study}, 
or due to the proximity effect~\cite{Zutic2018} with magnetic insulators~\cite{vobornik2011magnetic}, or ferromagnets~\cite{lv2018unidirectional,lee2018engineering}. 
We note, that the chiral perturbation of the electron spin density 
 manifests itself in various ways. 
For instance, probing the chiral spin textures induced on 
a surface 
by means of spin-polarized scanning tunneling microscopy~\cite{Wiesendanger2009} would be a new tool 
to access the parameters of the electron gas.
The chiral spin pattern 
in the electron gas can also induce a chiral
order of magnetic ions located either in the same material or 
in a different layer of a heterostructure due to proximity effect. 
Therefore, the phenomenon opens a way to record the information using magnetic skyrmions or similar chiral spin textures by electrical means.
Finally, the topological Hall effect is 
generally expected in
magnetic systems with spin-orbit interaction due to  asymmetric scattering of electrons on chiral spin textures~\cite{denisov2018general} pinned to defects and other inhomogeneities.
In particular, the considered mechanism could be responsible for the recently observed topological Hall effect in TI and DMS~\cite{jiang2019crossover,THE_TI,AronzonRozh}. 


The work has been supported by the Russian Science Foundation
(Project 18-72-10111), the Russian Foundation of Basic Research (grant 18-02-00668), and 
the Academy of Finland (Grant No.~318500). 
K.S.D. and N.S.A. thank the Foundation for the Advancement of Theoretical Physics and
Mathematics “BASIS”. 


\bibliography{Skyrmion}

\begin{thebibliography}{50}
\expandafter\ifx\csname natexlab\endcsname\relax\def\natexlab#1{#1}\fi
\expandafter\ifx\csname bibnamefont\endcsname\relax
  \def\bibnamefont#1{#1}\fi
\expandafter\ifx\csname bibfnamefont\endcsname\relax
  \def\bibfnamefont#1{#1}\fi
\expandafter\ifx\csname citenamefont\endcsname\relax
  \def\citenamefont#1{#1}\fi
\expandafter\ifx\csname url\endcsname\relax
  \def\url#1{\texttt{#1}}\fi
\expandafter\ifx\csname urlprefix\endcsname\relax\def\urlprefix{URL }\fi
\providecommand{\bibinfo}[2]{#2}
\providecommand{\eprint}[2][]{\url{#2}}

\bibitem[{\citenamefont{Wen et~al.}(1989)\citenamefont{Wen, Wilczek, and
  Zee}}]{Chir_Wilczek}
\bibinfo{author}{\bibfnamefont{X.~G.} \bibnamefont{Wen}},
  \bibinfo{author}{\bibfnamefont{F.}~\bibnamefont{Wilczek}}, \bibnamefont{and}
  \bibinfo{author}{\bibfnamefont{A.}~\bibnamefont{Zee}},
  \bibinfo{journal}{Phys. Rev. B} \textbf{\bibinfo{volume}{39}},
  \bibinfo{pages}{11413} (\bibinfo{year}{1989}).

\bibitem[{\citenamefont{Chen et~al.}(2010{\natexlab{a}})\citenamefont{Chen, Gu,
  and Wen}}]{chen2010local}
\bibinfo{author}{\bibfnamefont{X.}~\bibnamefont{Chen}},
  \bibinfo{author}{\bibfnamefont{Z.-C.} \bibnamefont{Gu}}, \bibnamefont{and}
  \bibinfo{author}{\bibfnamefont{X.-G.} \bibnamefont{Wen}},
  \bibinfo{journal}{Physical review b} \textbf{\bibinfo{volume}{82}},
  \bibinfo{pages}{155138} (\bibinfo{year}{2010}{\natexlab{a}}).

\bibitem[{\citenamefont{Batista et~al.}(2016)\citenamefont{Batista, Lin,
  Hayami, and Kamiya}}]{batista2016frustration}
\bibinfo{author}{\bibfnamefont{C.~D.} \bibnamefont{Batista}},
  \bibinfo{author}{\bibfnamefont{S.-Z.} \bibnamefont{Lin}},
  \bibinfo{author}{\bibfnamefont{S.}~\bibnamefont{Hayami}}, \bibnamefont{and}
  \bibinfo{author}{\bibfnamefont{Y.}~\bibnamefont{Kamiya}},
  \bibinfo{journal}{Reports on Progress in Physics}
  \textbf{\bibinfo{volume}{79}}, \bibinfo{pages}{084504}
  (\bibinfo{year}{2016}).

\bibitem[{\citenamefont{Kallin and Berlinsky}(2016)}]{kallin2016chiral}
\bibinfo{author}{\bibfnamefont{C.}~\bibnamefont{Kallin}} \bibnamefont{and}
  \bibinfo{author}{\bibfnamefont{J.}~\bibnamefont{Berlinsky}},
  \bibinfo{journal}{Reports on Progress in Physics}
  \textbf{\bibinfo{volume}{79}}, \bibinfo{pages}{054502}
  (\bibinfo{year}{2016}).

\bibitem[{\citenamefont{Kalmeyer and Laughlin}(1987)}]{Kalmeyer1987}
\bibinfo{author}{\bibfnamefont{V.}~\bibnamefont{Kalmeyer}} \bibnamefont{and}
  \bibinfo{author}{\bibfnamefont{R.~B.} \bibnamefont{Laughlin}},
  \bibinfo{journal}{Phys. Rev. Lett.} \textbf{\bibinfo{volume}{59}},
  \bibinfo{pages}{2095} (\bibinfo{year}{1987}).

\bibitem[{\citenamefont{Yang et~al.}(1993)\citenamefont{Yang, Warman, and
  Girvin}}]{yang1993possible}
\bibinfo{author}{\bibfnamefont{K.}~\bibnamefont{Yang}},
  \bibinfo{author}{\bibfnamefont{L.}~\bibnamefont{Warman}}, \bibnamefont{and}
  \bibinfo{author}{\bibfnamefont{S.}~\bibnamefont{Girvin}},
  \bibinfo{journal}{Physical review letters} \textbf{\bibinfo{volume}{70}},
  \bibinfo{pages}{2641} (\bibinfo{year}{1993}).

\bibitem[{\citenamefont{Kitaev}(2006)}]{kitaev2006anyons}
\bibinfo{author}{\bibfnamefont{A.}~\bibnamefont{Kitaev}},
  \bibinfo{journal}{Annals of Physics} \textbf{\bibinfo{volume}{321}},
  \bibinfo{pages}{2} (\bibinfo{year}{2006}).

\bibitem[{\citenamefont{Volovik}(2016)}]{Volovik2016}
\bibinfo{author}{\bibfnamefont{G.}~\bibnamefont{Volovik}},
  \bibinfo{journal}{JETP lett} \textbf{\bibinfo{volume}{103}},
  \bibinfo{pages}{140} (\bibinfo{year}{2016}).

\bibitem[{\citenamefont{Lee and Nagaosa}(1992)}]{lee1992gauge}
\bibinfo{author}{\bibfnamefont{P.~A.} \bibnamefont{Lee}} \bibnamefont{and}
  \bibinfo{author}{\bibfnamefont{N.}~\bibnamefont{Nagaosa}},
  \bibinfo{journal}{Physical Review B} \textbf{\bibinfo{volume}{46}},
  \bibinfo{pages}{5621} (\bibinfo{year}{1992}).

\bibitem[{\citenamefont{Bauer et~al.}(2014)\citenamefont{Bauer, Cincio, Keller,
  Dolfi, Vidal, Trebst, and Ludwig}}]{bauer2014chiral}
\bibinfo{author}{\bibfnamefont{B.}~\bibnamefont{Bauer}},
  \bibinfo{author}{\bibfnamefont{L.}~\bibnamefont{Cincio}},
  \bibinfo{author}{\bibfnamefont{B.~P.} \bibnamefont{Keller}},
  \bibinfo{author}{\bibfnamefont{M.}~\bibnamefont{Dolfi}},
  \bibinfo{author}{\bibfnamefont{G.}~\bibnamefont{Vidal}},
  \bibinfo{author}{\bibfnamefont{S.}~\bibnamefont{Trebst}}, \bibnamefont{and}
  \bibinfo{author}{\bibfnamefont{A.~W.} \bibnamefont{Ludwig}},
  \bibinfo{journal}{Nature communications} \textbf{\bibinfo{volume}{5}},
  \bibinfo{pages}{5137} (\bibinfo{year}{2014}).

\bibitem[{\citenamefont{Messio et~al.}(2017)\citenamefont{Messio, Bieri,
  Lhuillier, and Bernu}}]{MessioPRL2017}
\bibinfo{author}{\bibfnamefont{L.}~\bibnamefont{Messio}},
  \bibinfo{author}{\bibfnamefont{S.}~\bibnamefont{Bieri}},
  \bibinfo{author}{\bibfnamefont{C.}~\bibnamefont{Lhuillier}},
  \bibnamefont{and} \bibinfo{author}{\bibfnamefont{B.}~\bibnamefont{Bernu}},
  \bibinfo{journal}{Phys. Rev. Lett.} \textbf{\bibinfo{volume}{118}},
  \bibinfo{pages}{267201} (\bibinfo{year}{2017}).

\bibitem[{\citenamefont{Ye et~al.}(1999)\citenamefont{Ye, Kim, Millis,
  Shraiman, Majumdar, and Te\ifmmode \check{s}\else
  \v{s}\fi{}anovi\ifmmode~\acute{c}\else \'{c}\fi{}}}]{Ye1999}
\bibinfo{author}{\bibfnamefont{J.}~\bibnamefont{Ye}},
  \bibinfo{author}{\bibfnamefont{Y.~B.} \bibnamefont{Kim}},
  \bibinfo{author}{\bibfnamefont{A.~J.} \bibnamefont{Millis}},
  \bibinfo{author}{\bibfnamefont{B.~I.} \bibnamefont{Shraiman}},
  \bibinfo{author}{\bibfnamefont{P.}~\bibnamefont{Majumdar}}, \bibnamefont{and}
  \bibinfo{author}{\bibfnamefont{Z.}~\bibnamefont{Te\ifmmode \check{s}\else
  \v{s}\fi{}anovi\ifmmode~\acute{c}\else \'{c}\fi{}}}, \bibinfo{journal}{Phys.
  Rev. Lett.} \textbf{\bibinfo{volume}{83}}, \bibinfo{pages}{3737}
  (\bibinfo{year}{1999}).

\bibitem[{\citenamefont{Chun et~al.}(2000)\citenamefont{Chun, Salamon,
  Lyanda-Geller, Goldbart, and Han}}]{chun2000magnetotransport}
\bibinfo{author}{\bibfnamefont{S.}~\bibnamefont{Chun}},
  \bibinfo{author}{\bibfnamefont{M.}~\bibnamefont{Salamon}},
  \bibinfo{author}{\bibfnamefont{Y.}~\bibnamefont{Lyanda-Geller}},
  \bibinfo{author}{\bibfnamefont{P.}~\bibnamefont{Goldbart}}, \bibnamefont{and}
  \bibinfo{author}{\bibfnamefont{P.}~\bibnamefont{Han}},
  \bibinfo{journal}{Physical review letters} \textbf{\bibinfo{volume}{84}},
  \bibinfo{pages}{757} (\bibinfo{year}{2000}).

\bibitem[{\citenamefont{Bruno et~al.}(2004)\citenamefont{Bruno, Dugaev, and
  Taillefumier}}]{BrunoDugaev}
\bibinfo{author}{\bibfnamefont{P.}~\bibnamefont{Bruno}},
  \bibinfo{author}{\bibfnamefont{V.~K.} \bibnamefont{Dugaev}},
  \bibnamefont{and}
  \bibinfo{author}{\bibfnamefont{M.}~\bibnamefont{Taillefumier}},
  \bibinfo{journal}{Phys. Rev. Lett.} \textbf{\bibinfo{volume}{93}},
  \bibinfo{pages}{096806} (\bibinfo{year}{2004}).

\bibitem[{\citenamefont{Denisov et~al.}(2018)\citenamefont{Denisov, Rozhansky,
  Averkiev, and L{\"a}hderanta}}]{denisov2018general}
\bibinfo{author}{\bibfnamefont{K.}~\bibnamefont{Denisov}},
  \bibinfo{author}{\bibfnamefont{I.}~\bibnamefont{Rozhansky}},
  \bibinfo{author}{\bibfnamefont{N.}~\bibnamefont{Averkiev}}, \bibnamefont{and}
  \bibinfo{author}{\bibfnamefont{E.}~\bibnamefont{L{\"a}hderanta}},
  \bibinfo{journal}{Physical Review B} \textbf{\bibinfo{volume}{98}},
  \bibinfo{pages}{195439} (\bibinfo{year}{2018}).

\bibitem[{\citenamefont{Hamamoto et~al.}(2015)\citenamefont{Hamamoto, Ezawa,
  and Nagaosa}}]{QTHE}
\bibinfo{author}{\bibfnamefont{K.}~\bibnamefont{Hamamoto}},
  \bibinfo{author}{\bibfnamefont{M.}~\bibnamefont{Ezawa}}, \bibnamefont{and}
  \bibinfo{author}{\bibfnamefont{N.}~\bibnamefont{Nagaosa}},
  \bibinfo{journal}{Phys. Rev. B} \textbf{\bibinfo{volume}{92}},
  \bibinfo{pages}{115417} (\bibinfo{year}{2015}).

\bibitem[{\citenamefont{Fert et~al.}(2017)\citenamefont{Fert, Reyren, and
  Cros}}]{fert2017magnetic}
\bibinfo{author}{\bibfnamefont{A.}~\bibnamefont{Fert}},
  \bibinfo{author}{\bibfnamefont{N.}~\bibnamefont{Reyren}}, \bibnamefont{and}
  \bibinfo{author}{\bibfnamefont{V.}~\bibnamefont{Cros}},
  \bibinfo{journal}{Nature Reviews Materials} \textbf{\bibinfo{volume}{2}},
  \bibinfo{pages}{17031} (\bibinfo{year}{2017}).

\bibitem[{\citenamefont{Wiesendanger}(2016)}]{wiesendanger2016nanoscale}
\bibinfo{author}{\bibfnamefont{R.}~\bibnamefont{Wiesendanger}},
  \bibinfo{journal}{Nature Reviews Materials} \textbf{\bibinfo{volume}{1}},
  \bibinfo{pages}{16044} (\bibinfo{year}{2016}).

\bibitem[{\citenamefont{Soumyanarayanan
  et~al.}(2017)\citenamefont{Soumyanarayanan, Raju, Oyarce, Tan, Im,
  Petrovi{\'c}, Ho, Khoo, Tran, Gan et~al.}}]{soumyanarayanan2017tunable}
\bibinfo{author}{\bibfnamefont{A.}~\bibnamefont{Soumyanarayanan}},
  \bibinfo{author}{\bibfnamefont{M.}~\bibnamefont{Raju}},
  \bibinfo{author}{\bibfnamefont{A.~G.} \bibnamefont{Oyarce}},
  \bibinfo{author}{\bibfnamefont{A.~K.} \bibnamefont{Tan}},
  \bibinfo{author}{\bibfnamefont{M.-Y.} \bibnamefont{Im}},
  \bibinfo{author}{\bibfnamefont{A.~P.} \bibnamefont{Petrovi{\'c}}},
  \bibinfo{author}{\bibfnamefont{P.}~\bibnamefont{Ho}},
  \bibinfo{author}{\bibfnamefont{K.}~\bibnamefont{Khoo}},
  \bibinfo{author}{\bibfnamefont{M.}~\bibnamefont{Tran}},
  \bibinfo{author}{\bibfnamefont{C.}~\bibnamefont{Gan}}, \bibnamefont{et~al.},
  \bibinfo{journal}{Nature materials} \textbf{\bibinfo{volume}{16}},
  \bibinfo{pages}{898} (\bibinfo{year}{2017}).

\bibitem[{\citenamefont{Nakajima et~al.}(2017)\citenamefont{Nakajima, Oike,
  Kikkawa, Gilbert, Booth, Kakurai, Taguchi, Tokura, Kagawa, and
  Arima}}]{nakajima2017skyrmion}
\bibinfo{author}{\bibfnamefont{T.}~\bibnamefont{Nakajima}},
  \bibinfo{author}{\bibfnamefont{H.}~\bibnamefont{Oike}},
  \bibinfo{author}{\bibfnamefont{A.}~\bibnamefont{Kikkawa}},
  \bibinfo{author}{\bibfnamefont{E.~P.} \bibnamefont{Gilbert}},
  \bibinfo{author}{\bibfnamefont{N.}~\bibnamefont{Booth}},
  \bibinfo{author}{\bibfnamefont{K.}~\bibnamefont{Kakurai}},
  \bibinfo{author}{\bibfnamefont{Y.}~\bibnamefont{Taguchi}},
  \bibinfo{author}{\bibfnamefont{Y.}~\bibnamefont{Tokura}},
  \bibinfo{author}{\bibfnamefont{F.}~\bibnamefont{Kagawa}}, \bibnamefont{and}
  \bibinfo{author}{\bibfnamefont{T.-h.} \bibnamefont{Arima}},
  \bibinfo{journal}{Science advances} \textbf{\bibinfo{volume}{3}},
  \bibinfo{pages}{e1602562} (\bibinfo{year}{2017}).

\bibitem[{\citenamefont{Yu et~al.}(2018)\citenamefont{Yu, Koshibae, Tokunaga,
  Shibata, Taguchi, Nagaosa, and Tokura}}]{yu2018transformation}
\bibinfo{author}{\bibfnamefont{X.}~\bibnamefont{Yu}},
  \bibinfo{author}{\bibfnamefont{W.}~\bibnamefont{Koshibae}},
  \bibinfo{author}{\bibfnamefont{Y.}~\bibnamefont{Tokunaga}},
  \bibinfo{author}{\bibfnamefont{K.}~\bibnamefont{Shibata}},
  \bibinfo{author}{\bibfnamefont{Y.}~\bibnamefont{Taguchi}},
  \bibinfo{author}{\bibfnamefont{N.}~\bibnamefont{Nagaosa}}, \bibnamefont{and}
  \bibinfo{author}{\bibfnamefont{Y.}~\bibnamefont{Tokura}},
  \bibinfo{journal}{Nature} \textbf{\bibinfo{volume}{564}}, \bibinfo{pages}{95}
  (\bibinfo{year}{2018}).

\bibitem[{\citenamefont{Tokura et~al.}(2019)\citenamefont{Tokura, Yasuda, and
  Tsukazaki}}]{tokura2019magnetic}
\bibinfo{author}{\bibfnamefont{Y.}~\bibnamefont{Tokura}},
  \bibinfo{author}{\bibfnamefont{K.}~\bibnamefont{Yasuda}}, \bibnamefont{and}
  \bibinfo{author}{\bibfnamefont{A.}~\bibnamefont{Tsukazaki}},
  \bibinfo{journal}{Nature Reviews Physics} \textbf{\bibinfo{volume}{1}},
  \bibinfo{pages}{126} (\bibinfo{year}{2019}).

\bibitem[{\citenamefont{Okada et~al.}(2011)\citenamefont{Okada, Dhital, Zhou,
  Huemiller, Lin, Basak, Bansil, Huang, Ding, Wang et~al.}}]{OkadaPRL2011}
\bibinfo{author}{\bibfnamefont{Y.}~\bibnamefont{Okada}},
  \bibinfo{author}{\bibfnamefont{C.}~\bibnamefont{Dhital}},
  \bibinfo{author}{\bibfnamefont{W.}~\bibnamefont{Zhou}},
  \bibinfo{author}{\bibfnamefont{E.~D.} \bibnamefont{Huemiller}},
  \bibinfo{author}{\bibfnamefont{H.}~\bibnamefont{Lin}},
  \bibinfo{author}{\bibfnamefont{S.}~\bibnamefont{Basak}},
  \bibinfo{author}{\bibfnamefont{A.}~\bibnamefont{Bansil}},
  \bibinfo{author}{\bibfnamefont{Y.-B.} \bibnamefont{Huang}},
  \bibinfo{author}{\bibfnamefont{H.}~\bibnamefont{Ding}},
  \bibinfo{author}{\bibfnamefont{Z.}~\bibnamefont{Wang}}, \bibnamefont{et~al.},
  \bibinfo{journal}{Phys. Rev. Lett.} \textbf{\bibinfo{volume}{106}},
  \bibinfo{pages}{206805} (\bibinfo{year}{2011}).

\bibitem[{\citenamefont{Wei et~al.}(2013)\citenamefont{Wei, Katmis, Assaf,
  Steinberg, Jarillo-Herrero, Heiman, and Moodera}}]{WeiPRL2013}
\bibinfo{author}{\bibfnamefont{P.}~\bibnamefont{Wei}},
  \bibinfo{author}{\bibfnamefont{F.}~\bibnamefont{Katmis}},
  \bibinfo{author}{\bibfnamefont{B.~A.} \bibnamefont{Assaf}},
  \bibinfo{author}{\bibfnamefont{H.}~\bibnamefont{Steinberg}},
  \bibinfo{author}{\bibfnamefont{P.}~\bibnamefont{Jarillo-Herrero}},
  \bibinfo{author}{\bibfnamefont{D.}~\bibnamefont{Heiman}}, \bibnamefont{and}
  \bibinfo{author}{\bibfnamefont{J.~S.} \bibnamefont{Moodera}},
  \bibinfo{journal}{Phys. Rev. Lett.} \textbf{\bibinfo{volume}{110}},
  \bibinfo{pages}{186807} (\bibinfo{year}{2013}).

\bibitem[{\citenamefont{Chen et~al.}(2010{\natexlab{b}})\citenamefont{Chen,
  Chu, Analytis, Liu, Igarashi, Kuo, Qi, Mo, Moore, Lu
  et~al.}}]{chen2010massive}
\bibinfo{author}{\bibfnamefont{Y.}~\bibnamefont{Chen}},
  \bibinfo{author}{\bibfnamefont{J.-H.} \bibnamefont{Chu}},
  \bibinfo{author}{\bibfnamefont{J.}~\bibnamefont{Analytis}},
  \bibinfo{author}{\bibfnamefont{Z.}~\bibnamefont{Liu}},
  \bibinfo{author}{\bibfnamefont{K.}~\bibnamefont{Igarashi}},
  \bibinfo{author}{\bibfnamefont{H.-H.} \bibnamefont{Kuo}},
  \bibinfo{author}{\bibfnamefont{X.}~\bibnamefont{Qi}},
  \bibinfo{author}{\bibfnamefont{S.-K.} \bibnamefont{Mo}},
  \bibinfo{author}{\bibfnamefont{R.}~\bibnamefont{Moore}},
  \bibinfo{author}{\bibfnamefont{D.}~\bibnamefont{Lu}}, \bibnamefont{et~al.},
  \bibinfo{journal}{Science} \textbf{\bibinfo{volume}{329}},
  \bibinfo{pages}{659} (\bibinfo{year}{2010}{\natexlab{b}}).

\bibitem[{\citenamefont{Miron et~al.}(2010)\citenamefont{Miron, Gaudin,
  Auffret, Rodmacq, Schuhl, Pizzini, Vogel, and
  Gambardella}}]{miron2010current}
\bibinfo{author}{\bibfnamefont{I.~M.} \bibnamefont{Miron}},
  \bibinfo{author}{\bibfnamefont{G.}~\bibnamefont{Gaudin}},
  \bibinfo{author}{\bibfnamefont{S.}~\bibnamefont{Auffret}},
  \bibinfo{author}{\bibfnamefont{B.}~\bibnamefont{Rodmacq}},
  \bibinfo{author}{\bibfnamefont{A.}~\bibnamefont{Schuhl}},
  \bibinfo{author}{\bibfnamefont{S.}~\bibnamefont{Pizzini}},
  \bibinfo{author}{\bibfnamefont{J.}~\bibnamefont{Vogel}}, \bibnamefont{and}
  \bibinfo{author}{\bibfnamefont{P.}~\bibnamefont{Gambardella}},
  \bibinfo{journal}{Nature materials} \textbf{\bibinfo{volume}{9}},
  \bibinfo{pages}{230} (\bibinfo{year}{2010}).

\bibitem[{\citenamefont{Miron et~al.}(2011)\citenamefont{Miron, Garello,
  Gaudin, Zermatten, Costache, Auffret, Bandiera, Rodmacq, Schuhl, and
  Gambardella}}]{miron2011perpendicular}
\bibinfo{author}{\bibfnamefont{I.~M.} \bibnamefont{Miron}},
  \bibinfo{author}{\bibfnamefont{K.}~\bibnamefont{Garello}},
  \bibinfo{author}{\bibfnamefont{G.}~\bibnamefont{Gaudin}},
  \bibinfo{author}{\bibfnamefont{P.-J.} \bibnamefont{Zermatten}},
  \bibinfo{author}{\bibfnamefont{M.~V.} \bibnamefont{Costache}},
  \bibinfo{author}{\bibfnamefont{S.}~\bibnamefont{Auffret}},
  \bibinfo{author}{\bibfnamefont{S.}~\bibnamefont{Bandiera}},
  \bibinfo{author}{\bibfnamefont{B.}~\bibnamefont{Rodmacq}},
  \bibinfo{author}{\bibfnamefont{A.}~\bibnamefont{Schuhl}}, \bibnamefont{and}
  \bibinfo{author}{\bibfnamefont{P.}~\bibnamefont{Gambardella}},
  \bibinfo{journal}{Nature} \textbf{\bibinfo{volume}{476}},
  \bibinfo{pages}{189} (\bibinfo{year}{2011}).

\bibitem[{\citenamefont{Manchon et~al.}(2015)\citenamefont{Manchon, Koo, Nitta,
  Frolov, and Duine}}]{manchon2015new}
\bibinfo{author}{\bibfnamefont{A.}~\bibnamefont{Manchon}},
  \bibinfo{author}{\bibfnamefont{H.~C.} \bibnamefont{Koo}},
  \bibinfo{author}{\bibfnamefont{J.}~\bibnamefont{Nitta}},
  \bibinfo{author}{\bibfnamefont{S.}~\bibnamefont{Frolov}}, \bibnamefont{and}
  \bibinfo{author}{\bibfnamefont{R.}~\bibnamefont{Duine}},
  \bibinfo{journal}{Nature materials} \textbf{\bibinfo{volume}{14}},
  \bibinfo{pages}{871} (\bibinfo{year}{2015}).

\bibitem[{\citenamefont{Zhou et~al.}(2018)\citenamefont{Zhou, Song, Liu, Luan,
  Wang, Sun, Jiang, Xiang, Chen, Du et~al.}}]{zhou2018observation}
\bibinfo{author}{\bibfnamefont{L.}~\bibnamefont{Zhou}},
  \bibinfo{author}{\bibfnamefont{H.}~\bibnamefont{Song}},
  \bibinfo{author}{\bibfnamefont{K.}~\bibnamefont{Liu}},
  \bibinfo{author}{\bibfnamefont{Z.}~\bibnamefont{Luan}},
  \bibinfo{author}{\bibfnamefont{P.}~\bibnamefont{Wang}},
  \bibinfo{author}{\bibfnamefont{L.}~\bibnamefont{Sun}},
  \bibinfo{author}{\bibfnamefont{S.}~\bibnamefont{Jiang}},
  \bibinfo{author}{\bibfnamefont{H.}~\bibnamefont{Xiang}},
  \bibinfo{author}{\bibfnamefont{Y.}~\bibnamefont{Chen}},
  \bibinfo{author}{\bibfnamefont{J.}~\bibnamefont{Du}}, \bibnamefont{et~al.},
  \bibinfo{journal}{Science advances} \textbf{\bibinfo{volume}{4}},
  \bibinfo{pages}{eaao3318} (\bibinfo{year}{2018}).

\bibitem[{\citenamefont{Liu et~al.}(2008)\citenamefont{Liu, Qi, Dai, Fang, and
  Zhang}}]{liu2008quantum}
\bibinfo{author}{\bibfnamefont{C.-X.} \bibnamefont{Liu}},
  \bibinfo{author}{\bibfnamefont{X.-L.} \bibnamefont{Qi}},
  \bibinfo{author}{\bibfnamefont{X.}~\bibnamefont{Dai}},
  \bibinfo{author}{\bibfnamefont{Z.}~\bibnamefont{Fang}}, \bibnamefont{and}
  \bibinfo{author}{\bibfnamefont{S.-C.} \bibnamefont{Zhang}},
  \bibinfo{journal}{Physical review letters} \textbf{\bibinfo{volume}{101}},
  \bibinfo{pages}{146802} (\bibinfo{year}{2008}).

\bibitem[{\citenamefont{Yu et~al.}(2010)\citenamefont{Yu, Zhang, Zhang, Zhang,
  Dai, and Fang}}]{yu2010quantized}
\bibinfo{author}{\bibfnamefont{R.}~\bibnamefont{Yu}},
  \bibinfo{author}{\bibfnamefont{W.}~\bibnamefont{Zhang}},
  \bibinfo{author}{\bibfnamefont{H.-J.} \bibnamefont{Zhang}},
  \bibinfo{author}{\bibfnamefont{S.-C.} \bibnamefont{Zhang}},
  \bibinfo{author}{\bibfnamefont{X.}~\bibnamefont{Dai}}, \bibnamefont{and}
  \bibinfo{author}{\bibfnamefont{Z.}~\bibnamefont{Fang}},
  \bibinfo{journal}{Science} \textbf{\bibinfo{volume}{329}},
  \bibinfo{pages}{61} (\bibinfo{year}{2010}).

\bibitem[{\citenamefont{Chernyshov et~al.}(2009)\citenamefont{Chernyshov,
  Overby, Liu, Furdyna, Lyanda-Geller, and Rokhinson}}]{chernyshov2009evidence}
\bibinfo{author}{\bibfnamefont{A.}~\bibnamefont{Chernyshov}},
  \bibinfo{author}{\bibfnamefont{M.}~\bibnamefont{Overby}},
  \bibinfo{author}{\bibfnamefont{X.}~\bibnamefont{Liu}},
  \bibinfo{author}{\bibfnamefont{J.~K.} \bibnamefont{Furdyna}},
  \bibinfo{author}{\bibfnamefont{Y.}~\bibnamefont{Lyanda-Geller}},
  \bibnamefont{and} \bibinfo{author}{\bibfnamefont{L.~P.}
  \bibnamefont{Rokhinson}}, \bibinfo{journal}{Nature Physics}
  \textbf{\bibinfo{volume}{5}}, \bibinfo{pages}{656} (\bibinfo{year}{2009}).

\bibitem[{\citenamefont{Novik et~al.}(2005)\citenamefont{Novik,
  Pfeuffer-Jeschke, Jungwirth, Latussek, Becker, Landwehr, Buhmann, and
  Molenkamp}}]{novik2005band}
\bibinfo{author}{\bibfnamefont{E.}~\bibnamefont{Novik}},
  \bibinfo{author}{\bibfnamefont{A.}~\bibnamefont{Pfeuffer-Jeschke}},
  \bibinfo{author}{\bibfnamefont{T.}~\bibnamefont{Jungwirth}},
  \bibinfo{author}{\bibfnamefont{V.}~\bibnamefont{Latussek}},
  \bibinfo{author}{\bibfnamefont{C.}~\bibnamefont{Becker}},
  \bibinfo{author}{\bibfnamefont{G.}~\bibnamefont{Landwehr}},
  \bibinfo{author}{\bibfnamefont{H.}~\bibnamefont{Buhmann}}, \bibnamefont{and}
  \bibinfo{author}{\bibfnamefont{L.}~\bibnamefont{Molenkamp}},
  \bibinfo{journal}{Physical Review B} \textbf{\bibinfo{volume}{72}},
  \bibinfo{pages}{035321} (\bibinfo{year}{2005}).

\bibitem[{\citenamefont{Jungwirth et~al.}(2014)\citenamefont{Jungwirth,
  Wunderlich, Nov{\'a}k, Olejnik, Gallagher, Campion, Edmonds, Rushforth,
  Ferguson, and N{\v{e}}mec}}]{jungwirth2014spin}
\bibinfo{author}{\bibfnamefont{T.}~\bibnamefont{Jungwirth}},
  \bibinfo{author}{\bibfnamefont{J.}~\bibnamefont{Wunderlich}},
  \bibinfo{author}{\bibfnamefont{V.}~\bibnamefont{Nov{\'a}k}},
  \bibinfo{author}{\bibfnamefont{K.}~\bibnamefont{Olejnik}},
  \bibinfo{author}{\bibfnamefont{B.}~\bibnamefont{Gallagher}},
  \bibinfo{author}{\bibfnamefont{R.}~\bibnamefont{Campion}},
  \bibinfo{author}{\bibfnamefont{K.}~\bibnamefont{Edmonds}},
  \bibinfo{author}{\bibfnamefont{A.}~\bibnamefont{Rushforth}},
  \bibinfo{author}{\bibfnamefont{A.}~\bibnamefont{Ferguson}}, \bibnamefont{and}
  \bibinfo{author}{\bibfnamefont{P.}~\bibnamefont{N{\v{e}}mec}},
  \bibinfo{journal}{Reviews of Modern Physics} \textbf{\bibinfo{volume}{86}},
  \bibinfo{pages}{855} (\bibinfo{year}{2014}).

\bibitem[{\citenamefont{Li et~al.}(2013)\citenamefont{Li, Wang, Doǧan, and
  Manchon}}]{li2013tailoring}
\bibinfo{author}{\bibfnamefont{H.}~\bibnamefont{Li}},
  \bibinfo{author}{\bibfnamefont{X.}~\bibnamefont{Wang}},
  \bibinfo{author}{\bibfnamefont{F.}~\bibnamefont{Doǧan}}, \bibnamefont{and}
  \bibinfo{author}{\bibfnamefont{A.}~\bibnamefont{Manchon}},
  \bibinfo{journal}{Applied Physics Letters} \textbf{\bibinfo{volume}{102}},
  \bibinfo{pages}{192411} (\bibinfo{year}{2013}).

\bibitem[{\citenamefont{Hwang and Das~Sarma}(2007)}]{GrapheneHwang}
\bibinfo{author}{\bibfnamefont{E.~H.} \bibnamefont{Hwang}} \bibnamefont{and}
  \bibinfo{author}{\bibfnamefont{S.}~\bibnamefont{Das~Sarma}},
  \bibinfo{journal}{Phys. Rev. B} \textbf{\bibinfo{volume}{75}},
  \bibinfo{pages}{205418} (\bibinfo{year}{2007}).

\bibitem[{\citenamefont{Wunsch et~al.}(2006)\citenamefont{Wunsch, Stauber,
  Sols, and Guinea}}]{wunsch2006dynamical}
\bibinfo{author}{\bibfnamefont{B.}~\bibnamefont{Wunsch}},
  \bibinfo{author}{\bibfnamefont{T.}~\bibnamefont{Stauber}},
  \bibinfo{author}{\bibfnamefont{F.}~\bibnamefont{Sols}}, \bibnamefont{and}
  \bibinfo{author}{\bibfnamefont{F.}~\bibnamefont{Guinea}},
  \bibinfo{journal}{New Journal of Physics} \textbf{\bibinfo{volume}{8}},
  \bibinfo{pages}{318} (\bibinfo{year}{2006}).

\bibitem[{\citenamefont{Zhu et~al.}(2011)\citenamefont{Zhu, Yao, Zhang, and
  Chang}}]{ZhuPRL2011}
\bibinfo{author}{\bibfnamefont{J.-J.} \bibnamefont{Zhu}},
  \bibinfo{author}{\bibfnamefont{D.-X.} \bibnamefont{Yao}},
  \bibinfo{author}{\bibfnamefont{S.-C.} \bibnamefont{Zhang}}, \bibnamefont{and}
  \bibinfo{author}{\bibfnamefont{K.}~\bibnamefont{Chang}},
  \bibinfo{journal}{Phys. Rev. Lett.} \textbf{\bibinfo{volume}{106}},
  \bibinfo{pages}{097201} (\bibinfo{year}{2011}).

\bibitem[{\citenamefont{Checkelsky et~al.}(2012)\citenamefont{Checkelsky, Ye,
  Onose, Iwasa, and Tokura}}]{checkelsky2012dirac}
\bibinfo{author}{\bibfnamefont{J.~G.} \bibnamefont{Checkelsky}},
  \bibinfo{author}{\bibfnamefont{J.}~\bibnamefont{Ye}},
  \bibinfo{author}{\bibfnamefont{Y.}~\bibnamefont{Onose}},
  \bibinfo{author}{\bibfnamefont{Y.}~\bibnamefont{Iwasa}}, \bibnamefont{and}
  \bibinfo{author}{\bibfnamefont{Y.}~\bibnamefont{Tokura}},
  \bibinfo{journal}{Nature Physics} \textbf{\bibinfo{volume}{8}},
  \bibinfo{pages}{729} (\bibinfo{year}{2012}).

\bibitem[{\citenamefont{Gaj and Kossut}(2010)}]{GajKos}
\bibinfo{author}{\bibfnamefont{J.~A.} \bibnamefont{Gaj}} \bibnamefont{and}
  \bibinfo{author}{\bibfnamefont{J.}~\bibnamefont{Kossut}},
  \textbf{\bibinfo{volume}{144}}, \bibinfo{pages}{27} (\bibinfo{year}{2010}).

\bibitem[{\citenamefont{Kou et~al.}(2012)\citenamefont{Kou, Jiang, Lang, Xiu,
  He, Wang, Wang, Yu, Fedorov, Zhang et~al.}}]{kou2012magnetically}
\bibinfo{author}{\bibfnamefont{X.}~\bibnamefont{Kou}},
  \bibinfo{author}{\bibfnamefont{W.}~\bibnamefont{Jiang}},
  \bibinfo{author}{\bibfnamefont{M.}~\bibnamefont{Lang}},
  \bibinfo{author}{\bibfnamefont{F.}~\bibnamefont{Xiu}},
  \bibinfo{author}{\bibfnamefont{L.}~\bibnamefont{He}},
  \bibinfo{author}{\bibfnamefont{Y.}~\bibnamefont{Wang}},
  \bibinfo{author}{\bibfnamefont{Y.}~\bibnamefont{Wang}},
  \bibinfo{author}{\bibfnamefont{X.}~\bibnamefont{Yu}},
  \bibinfo{author}{\bibfnamefont{A.}~\bibnamefont{Fedorov}},
  \bibinfo{author}{\bibfnamefont{P.}~\bibnamefont{Zhang}},
  \bibnamefont{et~al.}, \bibinfo{journal}{Journal of Applied Physics}
  \textbf{\bibinfo{volume}{112}}, \bibinfo{pages}{063912}
  (\bibinfo{year}{2012}).

\bibitem[{\citenamefont{Watson et~al.}(2013)\citenamefont{Watson,
  Collins-McIntyre, Shelford, Coldea, Prabhakaran, Speller, Mousavi, Grovenor,
  Salman, Giblin et~al.}}]{watson2013study}
\bibinfo{author}{\bibfnamefont{M.}~\bibnamefont{Watson}},
  \bibinfo{author}{\bibfnamefont{L.}~\bibnamefont{Collins-McIntyre}},
  \bibinfo{author}{\bibfnamefont{L.}~\bibnamefont{Shelford}},
  \bibinfo{author}{\bibfnamefont{A.~I.} \bibnamefont{Coldea}},
  \bibinfo{author}{\bibfnamefont{D.}~\bibnamefont{Prabhakaran}},
  \bibinfo{author}{\bibfnamefont{S.}~\bibnamefont{Speller}},
  \bibinfo{author}{\bibfnamefont{T.}~\bibnamefont{Mousavi}},
  \bibinfo{author}{\bibfnamefont{C.~R.~M.} \bibnamefont{Grovenor}},
  \bibinfo{author}{\bibfnamefont{Z.}~\bibnamefont{Salman}},
  \bibinfo{author}{\bibfnamefont{S.}~\bibnamefont{Giblin}},
  \bibnamefont{et~al.}, \bibinfo{journal}{New Journal of Physics}
  \textbf{\bibinfo{volume}{15}}, \bibinfo{pages}{103016}
  (\bibinfo{year}{2013}).

\bibitem[{\citenamefont{{\v{Z}}uti{\'c}
  et~al.}(2018)\citenamefont{{\v{Z}}uti{\'c}, Matos-Abiague, Scharf, Dery, and
  Belashchenko}}]{Zutic2018}
\bibinfo{author}{\bibfnamefont{I.}~\bibnamefont{{\v{Z}}uti{\'c}}},
  \bibinfo{author}{\bibfnamefont{A.}~\bibnamefont{Matos-Abiague}},
  \bibinfo{author}{\bibfnamefont{B.}~\bibnamefont{Scharf}},
  \bibinfo{author}{\bibfnamefont{H.}~\bibnamefont{Dery}}, \bibnamefont{and}
  \bibinfo{author}{\bibfnamefont{K.}~\bibnamefont{Belashchenko}},
  \bibinfo{journal}{Materials Today}  (\bibinfo{year}{2018}).

\bibitem[{\citenamefont{Vobornik et~al.}(2011)\citenamefont{Vobornik, Manju,
  Fujii, Borgatti, Torelli, Krizmancic, Hor, Cava, and
  Panaccione}}]{vobornik2011magnetic}
\bibinfo{author}{\bibfnamefont{I.}~\bibnamefont{Vobornik}},
  \bibinfo{author}{\bibfnamefont{U.}~\bibnamefont{Manju}},
  \bibinfo{author}{\bibfnamefont{J.}~\bibnamefont{Fujii}},
  \bibinfo{author}{\bibfnamefont{F.}~\bibnamefont{Borgatti}},
  \bibinfo{author}{\bibfnamefont{P.}~\bibnamefont{Torelli}},
  \bibinfo{author}{\bibfnamefont{D.}~\bibnamefont{Krizmancic}},
  \bibinfo{author}{\bibfnamefont{Y.~S.} \bibnamefont{Hor}},
  \bibinfo{author}{\bibfnamefont{R.~J.} \bibnamefont{Cava}}, \bibnamefont{and}
  \bibinfo{author}{\bibfnamefont{G.}~\bibnamefont{Panaccione}},
  \bibinfo{journal}{Nano letters} \textbf{\bibinfo{volume}{11}},
  \bibinfo{pages}{4079} (\bibinfo{year}{2011}).

\bibitem[{\citenamefont{Lv et~al.}(2018)\citenamefont{Lv, Kally, Zhang, Lee,
  Jamali, Samarth, and Wang}}]{lv2018unidirectional}
\bibinfo{author}{\bibfnamefont{Y.}~\bibnamefont{Lv}},
  \bibinfo{author}{\bibfnamefont{J.}~\bibnamefont{Kally}},
  \bibinfo{author}{\bibfnamefont{D.}~\bibnamefont{Zhang}},
  \bibinfo{author}{\bibfnamefont{J.~S.} \bibnamefont{Lee}},
  \bibinfo{author}{\bibfnamefont{M.}~\bibnamefont{Jamali}},
  \bibinfo{author}{\bibfnamefont{N.}~\bibnamefont{Samarth}}, \bibnamefont{and}
  \bibinfo{author}{\bibfnamefont{J.-P.} \bibnamefont{Wang}},
  \bibinfo{journal}{Nature communications} \textbf{\bibinfo{volume}{9}},
  \bibinfo{pages}{111} (\bibinfo{year}{2018}).

\bibitem[{\citenamefont{Lee et~al.}(2018)\citenamefont{Lee, Richardella,
  Fraleigh, Liu, Zhao, and Samarth}}]{lee2018engineering}
\bibinfo{author}{\bibfnamefont{J.~S.} \bibnamefont{Lee}},
  \bibinfo{author}{\bibfnamefont{A.}~\bibnamefont{Richardella}},
  \bibinfo{author}{\bibfnamefont{R.~D.} \bibnamefont{Fraleigh}},
  \bibinfo{author}{\bibfnamefont{C.-x.} \bibnamefont{Liu}},
  \bibinfo{author}{\bibfnamefont{W.}~\bibnamefont{Zhao}}, \bibnamefont{and}
  \bibinfo{author}{\bibfnamefont{N.}~\bibnamefont{Samarth}},
  \bibinfo{journal}{npj Quantum Materials} \textbf{\bibinfo{volume}{3}},
  \bibinfo{pages}{51} (\bibinfo{year}{2018}).

\bibitem[{\citenamefont{Wiesendanger}(2009)}]{Wiesendanger2009}
\bibinfo{author}{\bibfnamefont{R.}~\bibnamefont{Wiesendanger}},
  \bibinfo{journal}{Rev. Mod. Phys.} \textbf{\bibinfo{volume}{81}},
  \bibinfo{pages}{1495} (\bibinfo{year}{2009}).

\bibitem[{\citenamefont{Jiang et~al.}(2019)\citenamefont{Jiang, Xiao, Wang,
  Shin, Andreoli, Zhang, Xiao, Zhao, Kayyalha, Zhang
  et~al.}}]{jiang2019crossover}
\bibinfo{author}{\bibfnamefont{J.}~\bibnamefont{Jiang}},
  \bibinfo{author}{\bibfnamefont{D.}~\bibnamefont{Xiao}},
  \bibinfo{author}{\bibfnamefont{F.}~\bibnamefont{Wang}},
  \bibinfo{author}{\bibfnamefont{J.-H.} \bibnamefont{Shin}},
  \bibinfo{author}{\bibfnamefont{D.}~\bibnamefont{Andreoli}},
  \bibinfo{author}{\bibfnamefont{J.}~\bibnamefont{Zhang}},
  \bibinfo{author}{\bibfnamefont{R.}~\bibnamefont{Xiao}},
  \bibinfo{author}{\bibfnamefont{Y.-F.} \bibnamefont{Zhao}},
  \bibinfo{author}{\bibfnamefont{M.}~\bibnamefont{Kayyalha}},
  \bibinfo{author}{\bibfnamefont{L.}~\bibnamefont{Zhang}},
  \bibnamefont{et~al.}, \bibinfo{journal}{arXiv preprint arXiv:1901.07611}
  (\bibinfo{year}{2019}).

\bibitem[{\citenamefont{Liu et~al.}(2017)\citenamefont{Liu, Zang, Ruan, Gong,
  He, Ma, Xue, and Wang}}]{THE_TI}
\bibinfo{author}{\bibfnamefont{C.}~\bibnamefont{Liu}},
  \bibinfo{author}{\bibfnamefont{Y.}~\bibnamefont{Zang}},
  \bibinfo{author}{\bibfnamefont{W.}~\bibnamefont{Ruan}},
  \bibinfo{author}{\bibfnamefont{Y.}~\bibnamefont{Gong}},
  \bibinfo{author}{\bibfnamefont{K.}~\bibnamefont{He}},
  \bibinfo{author}{\bibfnamefont{X.}~\bibnamefont{Ma}},
  \bibinfo{author}{\bibfnamefont{Q.-K.} \bibnamefont{Xue}}, \bibnamefont{and}
  \bibinfo{author}{\bibfnamefont{Y.}~\bibnamefont{Wang}},
  \bibinfo{journal}{Phys. Rev. Lett.} \textbf{\bibinfo{volume}{119}},
  \bibinfo{pages}{176809} (\bibinfo{year}{2017}).

\bibitem[{\citenamefont{Oveshnikov et~al.}(2015)\citenamefont{Oveshnikov,
  Kulbachinskii, Davydov, Aronzon, Rozhansky, Averkiev, Kugel, and
  Tripathi}}]{AronzonRozh}
\bibinfo{author}{\bibfnamefont{L.~N.} \bibnamefont{Oveshnikov}},
  \bibinfo{author}{\bibfnamefont{V.~A.} \bibnamefont{Kulbachinskii}},
  \bibinfo{author}{\bibfnamefont{A.~B.} \bibnamefont{Davydov}},
  \bibinfo{author}{\bibfnamefont{B.~A.} \bibnamefont{Aronzon}},
  \bibinfo{author}{\bibfnamefont{I.~V.} \bibnamefont{Rozhansky}},
  \bibinfo{author}{\bibfnamefont{N.~S.} \bibnamefont{Averkiev}},
  \bibinfo{author}{\bibfnamefont{K.~I.} \bibnamefont{Kugel}}, \bibnamefont{and}
  \bibinfo{author}{\bibfnamefont{V.}~\bibnamefont{Tripathi}},
  \bibinfo{journal}{Scientific Reports} \textbf{\bibinfo{volume}{5}},
  \bibinfo{pages}{17158} (\bibinfo{year}{2015}).

\end{thebibliography}

\onecolumngrid

\newpage

\begin{samepage}

\begin{figure*}
	\includegraphics[width=1.\textwidth]{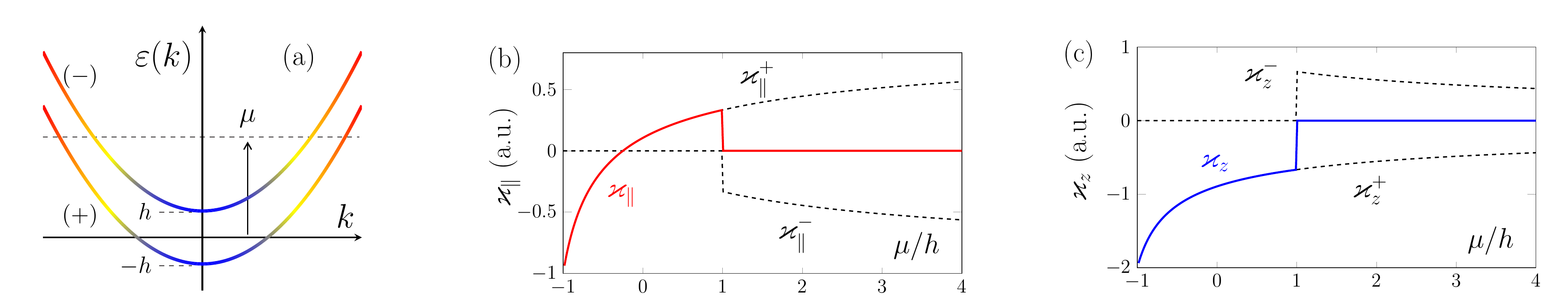}
	\caption{(a) Parabolic-like electron spectrum Eq.~\ref{eq_QW_spectr}, 
	(b,c) the dependence of $\varkappa_{\parallel,z}$ on $\mu$ for $\xi =0.5$.}	
	\label{f2}
\end{figure*}

\begin{figure*}
	\includegraphics[width=1.\textwidth]{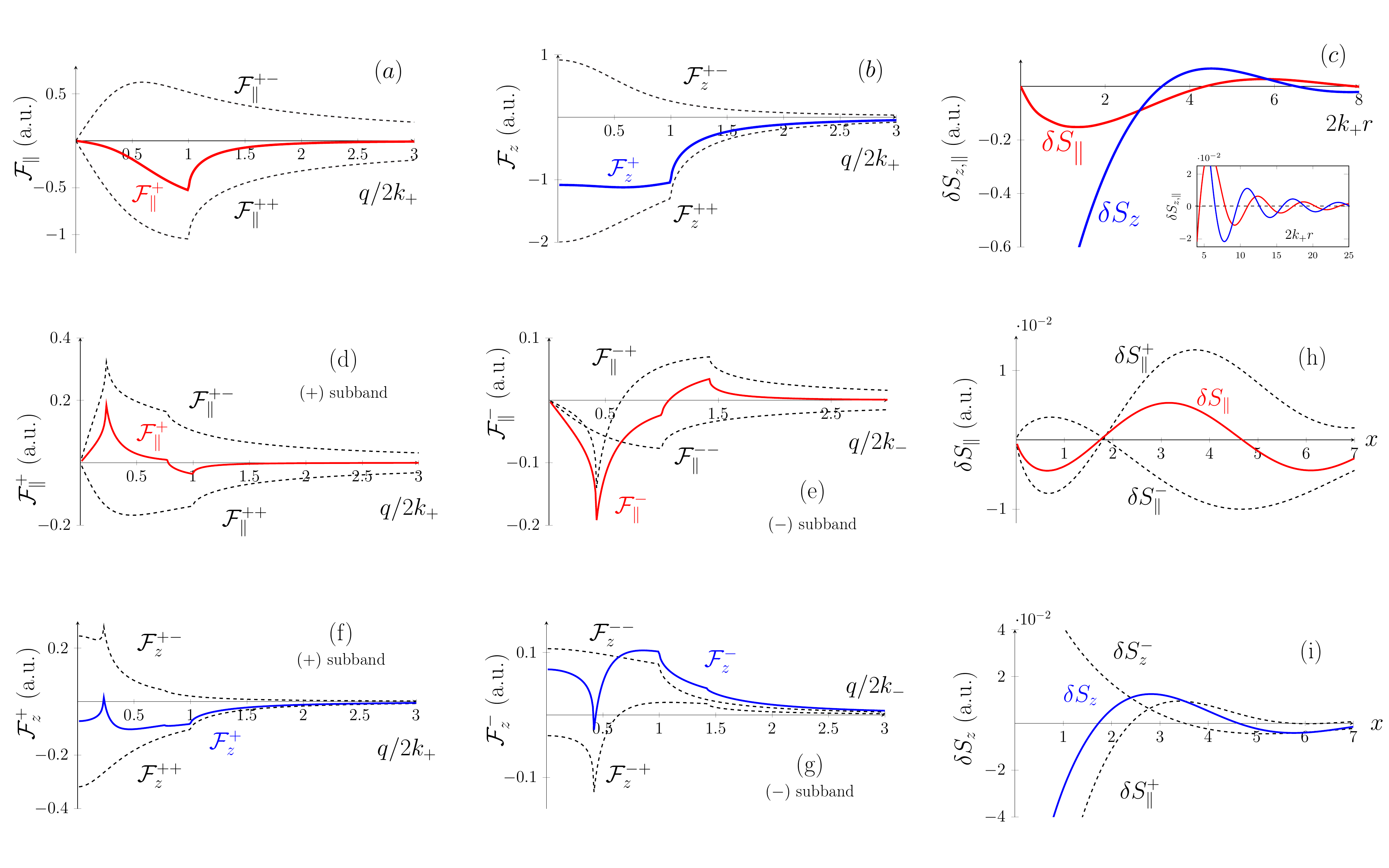}
	\caption{(a,b,c) The dependence of $\mathcal{F}_{z,\parallel}^+$ on $q/2k_+$, and $\delta S_{z,\parallel}$ on $2k_+r$ in case of one filled spin subband ($\xi=0.5$, $\mu=-0.4h$, $\zeta_+=1$), 
	(d-g) the dependence of $\mathcal{F}_{z,\parallel}^\pm$ on $q/2k_\pm$, 
	(h,i) the dependence of $\delta S_{z,\parallel}$ on $x=2 (\sqrt{2 m \mu}) r$ in case of two filled spin subbands, 
	the parameters are $\mu = 3.5h$, $\xi=0.5$, $\zeta_+ = 2.5$, $\zeta_- = 1.4$.}	
	\label{f3}
\end{figure*}

\begin{figure*}[h]
	\centering	
	\includegraphics[width=1.\textwidth]{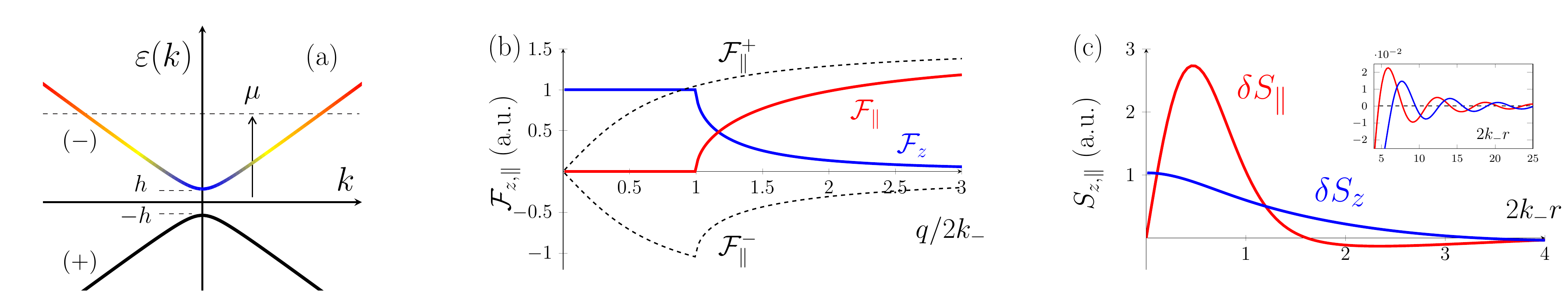}
	\caption{
	(a)  Dirac electron spectrum,  
	(b) the dependence of $\mathcal{F}_{z,\parallel}^\pm$ on $q/2k_-$, 
	(c) the dependence of $\delta S_{\parallel,z}$ on $2k_-r$, the inset shows the Friedel's oscillations. 
	The parameters are $\mu = 2h$, $\zeta_- = 1.7$, $ak_-=0.3$. }
	\label{f4}
\end{figure*}

\end{samepage}

\clearpage

\newpage

\section*{ONLINE SUPPORTING INFORMATION}

\appendix
\section{The spin-density response functions for the parabolic-like spectrum.}
\label{AppendixA}
Here we provide the derived analytical formulas for the spin-density response functions $\mathcal{F}_{z,\parallel}^\pm(q)$ in case of the electron parabolic-like spectrum $\varepsilon_k^\pm = k^2/2m \mp \sqrt{h^2 + (\lambda k)^2}$ (see Fig.~\ref{f2}a, Eq.~\ref{eq_QW_spectr} and the notation used in the main text). 
In the formulas below we use the following parameters:
$\xi = m\lambda^2/h<1$, $\zeta_\pm = \lambda k_\pm/h$, $k_\pm$ is the Fermi wavevector in the corresponding subband, $a_0 = \lambda/2h$ has a dimensionality of length. We find that $\mathcal{F}_{z,\parallel}^\pm(q)$ are decomposed onto the intra- and interband contributions as 
$\mathcal{F}_{z,\parallel}^{\pm} = \mathcal{F}_{z,\parallel}^{\pm \pm} + \mathcal{F}_{z,\parallel}^{\pm \mp}$. 
The intraband terms $\mathcal{F}_{z,\parallel}^{\pm \pm}(q)$ are given by:
\begin{align}
\label{eq_FparITRA}
    & \mathcal{F}_{\parallel}^{\pm \pm}(q) = \frac{m}{4\pi} \frac{1}{y(q)} \Bigl( \Theta[2k_{\pm}-q] \Phi_1(q) + \Theta[q-2k_{\pm}] \Phi_2^{\pm}(q) - \Phi_3^{\pm}(q)  \Bigr),
    \notag
    \\
    & y(q) = \sqrt{1+ (qa_0)^2 - \xi^2}, \qquad \Phi_1(q) = \ln{\sqrt{ 1 + (qa_0)^2 }}, 
  \qquad
    \Phi_3^{\pm}(q) = \tanh^{-1}\left( \frac{q a_0}{y(q)} \right) + \ln{\frac{y(q) \pm \xi (qa_0)}{\sqrt{1-\xi^2}}}, 
    \notag
    \\
    & \Phi_2^{\pm}(q) = \tanh^{-1}\left( \frac{a_0}{y(q)} \sqrt{q^2- 4k_{\pm}^2} \right) + \ln{\left[ y(q) \sqrt{1+\zeta_{\pm}^2} \pm \xi a_0 \sqrt{q^2- 4k_{\pm}^2} \right] } - \frac{1}{2}\ln{\left[ 1 + \zeta_{\pm}^2 - \xi^2 \right]}.
\end{align}
\begin{align}
\label{eq_FzITRA}
   & \mathcal{F}_z^{\pm \pm}(q)
= \mp \frac{m}{4 \pi} \frac{1}{y(q)} \frac{\xi}{q a_0} \Bigl( \Theta[2k_{\pm}-q] \Lambda_1^{\pm}(q) + \Theta[q-2 k_{\pm}] \Lambda_2^{\pm}(q) - \Lambda_3^{\pm}(q) \Bigr), \notag
\\
& \Lambda_1^{\pm}(q) =  y(q) \frac{\pi}{2 \xi} \mp \ln{ \sqrt{1+ (q a_0)^2}},
\hspace{0.4cm}
\Lambda_3^{\pm}(q) = \frac{y(q)}{\xi}  \tan^{-1}\left(\frac{1}{q a_0}\right) \mp \ln{\left[1+ q a_0 \frac{y(q) + q a_0}{1 \mp \xi}\right]}, 
\\
&\Lambda_2^{\pm}(q) = \frac{y(q)}{\xi}   \tan^{-1}\left( \frac{1}{a_0} \sqrt{ \frac{ 1+\zeta_{\pm}^2}{q^2 - 4k_{\pm}^2}}\right) \mp \left(  \ln{\left[ 1+ \left( q a_0\right)^2 + y(q) a_0 \sqrt{q^2- 4 k_{\pm}^2} \mp \xi \sqrt{1+\zeta_{\pm}^2}\right] } - \ln{\left[\sqrt{1+\zeta_{\pm}^2} \mp \xi \right]} \right),
\notag
\end{align}
and $\Theta[x]$ is the Heaviside function. 
The interband terms experience an additional symmetry $\mathcal{F}_z^{\pm \mp} =  \mathcal{F}_{\parallel}^{\pm \mp} (2m \lambda/q)$.
The functions $\mathcal{F}_{\parallel}^{\pm \mp}(q)$ are given by:
\begin{align}
\label{eq_Finter}
    & \mathcal{F}_{\parallel}^{+-}(q) = \frac{m}{4\pi} \frac{1}{y(q)} \times 
    \begin{cases}
 \mathcal{J}_+(q,1), &  \zeta_+ < 2 \sqrt{\xi + \xi^2}\\
\Theta[q_1^+ - q] \mathcal{J}_+(q,1) + \Theta[q - q_1^+] \Theta[q_2^+ - q] \mathcal{J}_+(q,x_+(q)) + \Theta[q- q_2^+] \mathcal{J}_+(q,1), &  \zeta_+ > 2 \sqrt{\xi + \xi^2}
\end{cases}
\notag
\\
&  \mathcal{F}_{\parallel}^{-+}(q) = \frac{m}{4\pi} \frac{1}{y(q)} 
 \Bigl( \Theta[q_1^- -q] \mathcal{J}_-(q,1) + \Theta[q-q_1^-] \Theta[q_2^--q] \mathcal{J}_-(q,|x_-(q)|)  + \Theta[q-q_2^-] \mathcal{J}_-(q,1) \Bigr),
\notag
\\
& \mathcal{J}_{\pm}(q,x) = \left[{\rm sgn}(q-q_0)\right]^{\frac{(1\mp 1)}{2}} \ln{
		\left[
		\frac{1 + \sqrt{F_{\pm}(q,0)}}{
			1 + \sqrt{F_{\pm}({q},x)}
		}
		\frac{\Delta_\pm(x)}{\Delta_\pm(0)}
		\right]
		}, 
		\qquad
		q_0 = \frac{1}{a_0} \sqrt{\xi + \xi^2},
\\
& 	F_{\pm}(q,x) = 1 + \xi \left(\frac{\Delta_\pm(x)}{y(q)}\right)^2 
	\left[ \left( \frac{2 k_\pm}{q} \right)^2- \left( \frac{\zeta_\pm}{\xi} \right)^2\right],
	\qquad
	\Delta_{\pm}(x) = \frac{\xi}{\zeta_\pm^2} \left[ \xi \mp \sqrt{1+ \zeta_\pm^2 x^2} \right], 
\notag
\\
& 	q_{1,2}^{\pm} = k_\pm {\left[(-1)^{1,2}\right]^{\frac{(1\mp 1)}{2}} } \left( 1 + (-1)^{1,2} \sqrt{1+ 4 \Delta_{\pm}(1)} \right),
\qquad 
	x_\pm(q) =  {\rm Re}\left[ \frac{q}{2k_\pm} \pm \frac{2m \lambda}{\zeta_\pm} \frac{y(q)}{\sqrt{q^2 - (2m\lambda)^2}} \right]. 
\notag
\end{align}

The dependence of $\mathcal{F}_{z,\parallel}^\pm(q)$ on $q$ is shown in Fig.~\ref{f3} and is thoughtfully discussed in the main text. In the limit of $q\to 0$ these functions behave as $\mathcal{F}_{z}^\pm(q) \approx \varkappa_{z}^\pm $, and 
$\mathcal{F}_{\parallel}^\pm(q) \approx q\cdot \varkappa_\parallel^\pm$, 
the coefficients $\varkappa_{z,\parallel}^\pm$ describing the local coupling regime are found to be:
\begin{align}
&\varkappa_{z}^\pm = \mp \frac{m}{4\pi} \frac{\Theta[\mu \pm h]}{\sqrt{1+\zeta_\pm^2} \mp \xi},
\qquad \varkappa_{\parallel}^\pm= \pm \frac{1}{8 \pi \lambda}\left( 1 - \frac{1}{\sqrt{1+\zeta_\pm^2} \mp \xi} \right) \Theta[\mu \pm  h].
\end{align}

\newpage
\section{Derivation of spin-density response functions.}

We consider a two-dimensional electron gas with an effective magnetic field acting on an electron spin in $k$-space: $\boldsymbol{B}_k = \left(    \lambda k \cos{\left(\chi \varphi_k+\gamma \right)}, \lambda k \sin{\left(\chi \varphi_k + \gamma \right)}, h\right)$, 
where $(\lambda,h)>0$, $\chi = \pm 1$, $\gamma$ is an arbitrary real number. 
There are two spin subbands $s=\pm$, an electron in state $(\boldsymbol{k},s)$
has its spin $\boldsymbol{S}_k^{\pm}$ parallel $(s=+)$ or antiparallel $(s=-)$ to $\boldsymbol{B}_k$. 
The corresponding spinors $|u_k^{s} \rangle$ for $(\bm{k},s)$ states are given:
\begin{align}
& |u_k^+ \rangle = \begin{pmatrix}
e^{-i(\chi \varphi_k + \gamma)} a_k\\
b_k
\end{pmatrix},
\hspace{1cm}
|u_k^- \rangle = \begin{pmatrix}
b_k \\
- e^{i (\chi \varphi_k + \gamma)} a_k
\end{pmatrix},
\\
& a_k = \sqrt{\frac{1+n_k}{2}} = \frac{h + B_k}{\sqrt{ (\lambda k)^2 + (h + B_k)^2 }}, 
\hspace{1cm}
b_k =\sqrt{\frac{1-n_k}{2}} = \frac{\lambda k}{\sqrt{ (\lambda k)^2 + (h + B_k)^2 }},
\notag
\end{align}
where $n_k = h/B_k$, and $B_k = \sqrt{h^2 + (\lambda k)^2}$. 
The static spin-density response functions $\mathcal{F}_\alpha(\bm q)$ introduced in the main text Eq.~\ref{eq_First}
contain contributions from each spin subband and are given by:
\begin{align}
    & \mathcal{F}_\alpha(\bm q) = \mathcal{F}_\alpha^+(\bm q) + \mathcal{F}_\alpha^-(\bm q), \qquad    
\mathcal{F}_{\alpha}^s(\boldsymbol{q}) = \sum_{k,s'=\pm} f_k^s \left( 
    \frac{\langle u_k^s| \hat{S}_{\alpha} | u_{k+q}^{s'} \rangle \langle u_{k+q}^{s'}| u_{k}^{s} \rangle}{\varepsilon_k^s - \varepsilon_{k+q}^{s'}+i0} + \frac{\langle u_{k-q}^{s'}| \hat{S}_{\alpha} | u_{k}^{s} \rangle \langle u_{k}^{s}| u_{k-q}^{s'} \rangle}{\varepsilon_k^s - \varepsilon_{k-q}^{s'}-i0}
    \right), 
\end{align}
where $\hat{S}_{\alpha} = \hat{\sigma}_{\alpha}/2$, 
$\hat{\sigma}_\alpha$ is the Pauli matrix, $\alpha = (x,y,z)$, 
$f_k^s$ and $\varepsilon_k^s$ are the distribution function and an electron energy in state $(\bm{k},s)$. 
Replacing $\boldsymbol{k} \to - \boldsymbol{k}$ in the last integral (we assume that $\varepsilon_k^s = \varepsilon_{-k}^s$) and using the following relations:
\begin{align}
    & \langle u_{-k-q}^{s'}| \hat{S}_{z} | u_{-k}^{s} \rangle \langle u_{-k}^{s}| u_{-k-q}^{s'} \rangle = 
    \langle u_{k+q}^{s'}| \hat{S}_{z} | u_{k}^{s} \rangle \langle u_{k}^{s}| u_{k+q}^{s'} \rangle,
    \\
    & \langle u_{-k-q}^{s'}| \hat{S}_{x,y} | u_{-k}^{s} \rangle \langle u_{-k}^{s}| u_{-k-q}^{s'} \rangle = - \langle u_{k+q}^{s'}| \hat{S}_{x,y} | u_{k}^{s} \rangle \langle u_{k}^{s}| u_{k+q}^{s'} \rangle, \notag
\end{align}
we get for the response functions at zero temperature:
\begin{align}
\label{eq_Fxyz}
&  \mathcal{F}_{\alpha}^s(\boldsymbol{q}) = \int\limits_0^{k_s}\frac{kdk}{2\pi} \mathcal{P} \int\limits_0^{2\pi}\frac{d\theta}{2\pi} \sum_{s'=\pm} 
    \frac{\mathcal{L}_{\alpha,kq}^{ss'} }{\varepsilon_k^s - \varepsilon_{k+q}^{s'}}, 
 \\
 & \mathcal{L}_{z,kq}^{ss'}  = {\rm Re}\left[ \langle u_k^s| \hat{\sigma}_{z} | u_{k+q}^{s'} \rangle \langle u_{k+q}^{s'}| u_{k}^{s} \rangle \right],
 \quad
 \mathcal{L}_{xy,kq}^{ss'}  = i \times {\rm Im}\left[ \langle u_k^s| \hat{\sigma}_{x,y} | u_{k+q}^{s'} \rangle \langle u_{k+q}^{s'}| u_{k}^{s} \rangle \right],
 \notag
\end{align}
where $\mathcal{P}$ stands for the principal value, $\theta$ is the polar angle between $\bm{k}$ and $\bm{q}$, $k_\pm$ is the Fermi wavevector in $s$ subband.  
We note that $\mathcal{L}_{xy,kq}^{ss'}$ are purely imaginary. 
It follows from the structure of $\mathcal{L}_{xy,kq}^{ss'}$, that the response functions $\mathcal{F}_{x,y}^s(\bm q)$ for the in-plane spin components can be presented in form:
\begin{align}
& \mathcal{L}_{xy,kq}^{ss'} = i \left(\hat{n} \bm{e}_q \right)_{x,y} \mathcal{L}_{\parallel,kq}^{ss'}, 
    \qquad 
    \mathcal{F}_{x,y}^s(\bm{q}) = i (\hat{n} \bm{e}_q)_{x,y} \mathcal{F}_\parallel^s(q), 
    \\
&   \mathcal{F}_\parallel^s(q) = \int\limits_0^{k_s}\frac{kdk}{2\pi} \mathcal{P} \int\limits_0^{2\pi}\frac{d\theta}{2\pi} 
\sum_{s'=\pm} 
    \frac{\mathcal{L}_{\parallel,kq}^{ss'} }{\varepsilon_k^s - \varepsilon_{k+q}^{s'}}, 
    \qquad
    \hat{n} = \begin{pmatrix}
    \sin{\gamma} & (-1)^\chi \cos{\gamma}
    \\ -\cos{\gamma} & (-1)^\chi \sin{\gamma}
    \end{pmatrix},
    \qquad 
\end{align}
where $\hat{n}$ is an orthogonal matrix determined by a particular spin-orbit interaction type, 
$\bm{e}_q = \bm{q}/q$ is the unit vector in the direction of $\bm q$, the functions 
$\mathcal{F}_\parallel^\pm(q)$ are real and depend only on $q$ modulus. 
The explicit expressions for the matrix elements $\mathcal{L}_{z,\parallel,kq}^{ss'}$ are given:
\begin{equation}
    \mathcal{L}_{z,kq}^{ss'},  = \frac{(-1)^{(1-s)\over 2}}{2} (n_k + (-1)^{s-s'}n_{k+q}),
 \qquad 
 \mathcal{L}_{\parallel,kq}^{ss'},  = (-1)^{s-s'} (q a_0) n_k n_{k+q},
 \qquad
 a_0 = \frac{\lambda}{2h}.
\end{equation}
When calculating these matrix elements the following relation is very useful: $(a_k b_k)/(a_k^2 - b_k^2) = k a_0$. 
The details of the spin-density response depend on a particular electron spectrum, below 
we provide the calculations for parabolic-like and Dirac types of spectra.

\subsection{Integrals for the Dirac spectrum}
Here we present the calculations of $\mathcal{F}_{z,\parallel}^\pm(q)$ for the Dirac electron spectrum $\varepsilon_k^{\pm} = \mp B_k = \mp \sqrt{h^2 + (\lambda k)^2}$, see Eq.~\ref{eq_TI_spectr} and Fig.\ref{f4}a in the main text. 
We write the spin-density response functions in form:
\begin{align}
\label{eqA-F}
&
\begin{pmatrix}
\mathcal{F}_{\parallel}^{s}(q)
\\
\mathcal{F}_{z}^{s}(q)
\end{pmatrix}
 = - \int\limits_0^{k_{s}} \frac{kdk}{2\pi} \mathcal{P} \int\limits_0^{2\pi}\frac{d\theta}{2\pi} 
 \begin{pmatrix}
 s \mathcal{A}_{k,q} \times (qa_0)
 \\
 \mathcal{B}_{k,q}
 \end{pmatrix},
\\
& \mathcal{A}_{k,q} =  \left[ \frac{1}{B_k - B_{k+q}} - \frac{1}{B_k + B_{k+q}} \right] n_k n_{k+q} = - \frac{1}{q}  \left( \frac{ m_g}{ B_k} \right) \frac{2h}{q + 2 k \cos{\theta}},
\notag
\\
& \mathcal{B}_{k,q} = \frac{1}{2} \left[ \frac{n_k + n_{k+q} }{B_k - B_{k+q}} + \frac{n_k - n_{k+q}}{B_k + B_{k+q}} \right]  = -  \frac{2h}{q + 2 k \cos{\theta}}, 
\notag
\end{align}
where we introduced the parameter $m_g = {h}/{\lambda^2}$, which is an effective mass at the bottom of subband $k\approx 0$. 
The integral over the angle in Eq.~\ref{eqA-F} for both $\mathcal{F}_{z,\parallel}^s$ is taken using:
\begin{equation}
\label{eq_Ap_angle}
\mathcal{P} \int\limits_0^{2\pi} \frac{d\theta}{2\pi} \frac{1}{a+ b\cos{\theta}} = \frac{\Theta[a-b]}{\sqrt{a^2-b^2}},
\qquad
a,b>0. 
\end{equation}
The remaining integrals over $k$ are taken using $I_z$ for $\mathcal{F}_{z}^s$, and $I_{\parallel}$ for $\mathcal{F}_{\parallel}^s$ correspondingly:
\begin{align}
\label{eq_IparZ}
& I_{z} = 
\int\limits_0^{1} {xdx} \frac{\Theta[z-x]}{\sqrt{z^2-x^2}} =   z - \Theta[z-1] \sqrt{z^2-1},
\\
& I_{\parallel} = 
\int\limits_0^{1} {xdx} \frac{1}{\sqrt{y^2 + x^2}} \frac{\Theta[z-x]}{\sqrt{z^2-x^2}} = \tan^{-1}\left(\frac{z}{y}\right) - \Theta[z-1] \tan^{-1}\left( \sqrt{ \frac{z^2-1}{y^2+1}} \right). \notag
\end{align}
Let us consider the case when the Fermi energy $\mu>h$ lies in the upper subband (see Fig.\ref{f4}a). 
The response from partially filled $(-)$ subband, using the integrals in Eq.~(\ref{eq_Ap_angle},\ref{eq_IparZ}), is given by:
\begin{align}
    & \mathcal{F}_z^{-}(q) = \frac{m_g}{4\pi} \left( 1 - \Theta[q-2k_{-}] \sqrt{1 - (q/2k_{-})^2} \right),
    \notag
    \\
    & \mathcal{F}_{\parallel}^{-}(q) = - \frac{m_g}{4\pi} \left[
    \tan^{-1}\left(qa_0\right) - \Theta[q-2k_{-}] 
    \tan^{-1}\left( a_0 \sqrt{ \frac{q^2-4k_{-}^2}{1+\zeta_-^2}} \right)
    \right],
\end{align}
here $\zeta_- = \lambda k_-/h$. 
Considering the response from the fully filled $(+)$ subband we should 
take the integrals Eq.~\ref{eq_IparZ} in the limit $k_+ \to \infty$. At that no response of spin $z$-component is induced ($\mathcal{F}_z^+ =0$), while the in-plane response function is given by: 
\begin{align}
    & \mathcal{F}_{\parallel}^{+}(q) = \frac{m_g}{4\pi}
    \tan^{-1}\left(qa_0\right).
\end{align}

\newpage

\subsection{Integrals for the parabolic-like spectrum}
Here we present the calculations of $\mathcal{F}_{z,\parallel}^\pm(q)$ for the parabolic-like electron spectrum $\varepsilon_k^{\pm} = k^2/2m \mp \sqrt{h^2 + (\lambda k)^2}$, see Eq.~\ref{eq_QW_spectr} and Fig.\ref{f2}a in the main text. 
We write the spin-density response functions in form:
\begin{align}
\label{eqA-F2}
&
\begin{pmatrix}
\mathcal{F}_{\parallel}^{s}(q)
\\
\mathcal{F}_{z}^{s}(q)
\end{pmatrix}
 = \int\limits_0^{k_{s}} \frac{kdk}{2\pi} \mathcal{P} \int\limits_0^{2\pi}\frac{d\theta}{2\pi} 
 \begin{pmatrix}
 \mathcal{C}_{k,q}^s \times (qa_0)
 \\
 \mathcal{D}_{k,q}^s
 \end{pmatrix},
\\
& \mathcal{C}_{k,q}^\pm =  \left[ \frac{1}{ (\delta \varepsilon_{k,q} \mp B_k) \pm B_{k+q}} - \frac{1}{(\delta \varepsilon_{k,q} \mp B_k) \mp B_{k+q}} \right] n_k n_{k+q} = \mp \frac{2h  n_k}{(\delta \varepsilon_{k,q} \mp B_k)^2 - B_{k+q}^2}
\notag
\\
& \mathcal{D}_{k,q}^\pm = \pm \frac{1}{2} \left[ \frac{n_k + n_{k+q} }{(\delta \varepsilon_{k,q} \mp B_k) \pm B_{k+q}} + \frac{n_k - n_{k+q}}{(\delta \varepsilon_{k,q} \mp B_k) \mp B_{k+q}} \right]  = \frac{ \pm n_k \delta \varepsilon_{k,q} - 2h  }{(\delta \varepsilon_{k,q} \mp B_k)^2 - B_{k+q}^2}, 
\notag
\\
& \delta \varepsilon_{k,q} = - \frac{q^2}{2m} - \frac{kq}{m} \cos{\theta}. 
\notag
\end{align}
The denominator in the $\mathcal{C}_{k,q}^\pm, \mathcal{D}_{k,q}^\pm$ can be expressed as $
(\delta \varepsilon_{k,q} \mp B_k)^2 - B_{k+q}^2 = (kq/m)^2 \times (a_\pm + b_\pm \cos{\theta} + \cos^2{\theta})
$, where the coefficients $a_\pm, b_\pm$ do not depend on $\theta$.  
There are two types of integrals with respect to the angle $\theta$:
\begin{align}
    & 
I_1^\pm =    \mathcal{P} \int\limits_0^{2\pi}\frac{d\theta}{2\pi} \frac{1}{a_\pm + b_\pm \cos{\theta} + \cos^2{\theta}} = 
    \pm \frac{x^2 t}{ | \Delta_\pm(x)|} 
    \left[ 
    \frac{1}{w(t,x)} - \frac{{\rm sgn}\left( u_\pm(x,t) \right) }{w(u_\pm,x)}
    \right],
    \\
 & I_2^\pm =  \mathcal{P} \int\limits_0^{2\pi}\frac{d\theta}{2\pi} \frac{\cos{\theta}}{a_\pm + b_\pm \cos{\theta} + \cos^2{\theta}} =
 \pm \frac{xt}{|\Delta_\pm(x)|}  
 \left[ 
 {\rm sgn}(u_\pm(x,t)) \frac{u_\pm(x,t)}{w(u_\pm,x)} - \frac{t}{w(t,x)}
 \right],
\end{align}
where we introduced the following notation: $x= k/k_\pm$, $t = q/2k_\pm$, the functions $\Delta_\pm(x), u_\pm(x,t), w(t,x)$ are given by:
\begin{align*}
& \Delta_\pm(x) = \frac{\xi}{\zeta_\pm^2} \left[ \xi \mp \sqrt{1+ x^2 \zeta_\pm^2} \right] \lessgtr 0, 
\quad  u_\pm(x,t) = t - \frac{\Delta_\pm(x)}{t},
\quad w(t,x) = {\rm Re} \left[ \sqrt{t^2-x^2} \right] = \Theta\left(|t|-x\right) \sqrt{t^2-x^2}, 
\end{align*}
here $\xi = m \lambda^2/h$, $\zeta_\pm = \lambda k_\pm /h$. 
For the calculation of $\mathcal{F}_\parallel^\pm$ we need only $I_1^\pm$ integrals, while calculating $\mathcal{F}_z^\pm$ requires both $I_{1,2}^\pm$. 
After the integration over the angle we can decompose the integrals over $k$ as 
$\mathcal{F}_{\parallel,z}^{\pm}(t) = \mathcal{F}_{\parallel,z}^{\pm \pm}(t) + \mathcal{F}_{\parallel,z}^{\pm \mp}(t)$, where 
\begin{align}
    & \mathcal{F}_{\parallel}^{\pm \pm}(t) = - Q_\pm \mathcal{I}_{1}^{\pm}, 
    \qquad 
\mathcal{F}_{z}^{\pm \pm}(t) = \mp Q_\pm \frac{1}{q a_0}  \mathcal{I}_{2}^{\pm},   \qquad
\mathcal{F}_\parallel^{\pm \mp}(t) = \left( \frac{q}{2m \lambda} \right) \mathcal{F}_z^{\pm \mp}(t) = Q_\pm  \mathcal{I}_{3}^{\pm}, 
\end{align}
here the pre-factor $Q_\pm = ({m}/{4\pi}) \left({m \lambda}/{k_\pm}\right)$, the integrals $\mathcal{I}_{1,2,3}^\pm$ are shown below: 
\begin{align}
    & 
    \mathcal{I}_1^\pm = \int\limits_0^{1} \frac{xdx}{|\Delta_{\pm}(x)|} \frac{\Theta[t-x]}{\sqrt{t^2-x^2}} \frac{1}{\sqrt{1 + \zeta_\pm^2 x^2} },
    \\
    &
    \mathcal{I}_2^\pm =  \int\limits_0^{1} \frac{xdx }{ |\Delta_{\pm}(x)| }  \frac{\Theta[t-x]}{\sqrt{t^2-x^2}},
    \\
    & \mathcal{I}_3^\pm = \int\limits_0^{1} \frac{xdx}{|\Delta_{\pm}(x)| } 
 \frac{\Theta\left[\left| 
 	u_\pm(x,t) \right| -x\right]}{\sqrt{u_\pm^2(x,t)-x^2}} 
 	 \frac{{\rm sgn}\left[u_\pm(x,t)\right]}{\sqrt{1 + \zeta_\pm^2 x^2} }. 
 \end{align}

The terms $\mathcal{F}_{\parallel,z}^{\pm \pm}$, $\mathcal{F}_{\parallel,z}^{\pm \mp}$ listed in the Appendix~\ref{AppendixA} are reproduced after some straightforward calculations of $\mathcal{I}_{1,2,3}^\pm$ given above. 
Let us only note the complex range of integration in $\mathcal{I}_3^{\pm}$, which leads to a double spike structure of the the interband response functions. This feature reflects the presence of two nesting vectors of the Fermi surface at the interband transitions.

\end{document}